\documentclass[12pt]{article}
\usepackage{graphicx}
\usepackage{hyperref}
\usepackage{amsmath}
\usepackage{amscd,amssymb}
\usepackage[cmtip,arrow,
matrix]{xy}
\usepackage[mathscr]{euscript}

\usepackage{latexsym}
\usepackage{graphics}
\usepackage{color}

\usepackage{graphicx}

\usepackage{curves}

\usepackage{graphicx}

\input{xy}
\xyoption{all}

\newcommand{\cE}{\mathcal{E}}

\newcommand{\cG}{\mathcal{G}}

\newcommand{\Homsf}{\mathsf{Hom}}
\newcommand{\sfs}{\mathsf{s}}
\newcommand{\sft}{\mathsf{t}}

\newcommand{\scrG}{\mathscr{G}}
\newcommand{\scrF}{\mathscr{F}}
\newcommand{\scrH}{\mathscr{H}}
\newcommand{\scrQ}{\mathscr{Q}}
\newcommand{\scrK}{\mathscr{K}}

\def\cS{{\mathcal S}}

\def\cM{{\mathcal M}}

\def\cB{{\mathcal B}}

\def\cK{{\mathcal K}}

\def\cH{{\mathcal H}}

\def\cT{{\mathcal T}}

\def\cE{{\mathcal E}}

\def\cX{{\mathcal X}}

\def\cG{{\mathcal G}}

\def\cU{{\mathcal U}}
\def\cV{{\mathcal V}}
\def\cW{{\mathcal W}}

\def\mD{{\mathfrak D}}

\def\frp{{\mathfrak p}}

\newtheorem{theorem}[equation]{Theorem}

\newtheorem{corollary}[equation]{Corollary}
\newtheorem{proposition}[equation]{Proposition}

\newtheorem{definition}[equation]{Definition}

\def\ii{{\,{\rm i}\,}}
\def\dd{{\rm d}}

\newcommand{\IZ}{\mathbb{Z}}
\newcommand{\IC}{\mathbb{C}}
\newcommand{\IP}{\mathbb{P}}

\newcommand{\IR}{\mathbb{R}}
\newcommand{\IQ}{\mathbb{Q}}

\newcommand{\beq}{\begin{eqnarray}}
\newcommand{\eeq}{\end{eqnarray}}
\numberwithin{equation}{section}

\begin{document}

\hfill EMPG--15--02
\vspace{0.5in}

\begin{center}
{\large\bf Stratified Fiber Bundles, Quinn Homology \\[10pt] and Brane Stability of Hyperbolic Orbifolds}
\end{center}

\vspace{0.1in}

\begin{center}
{\large
Andrey A. Bytsenko$^{(a),}$\footnote{aabyts@gmail.com}, \ 
Richard J. Szabo$^{(b),}$\footnote{R.J.Szabo@hw.ac.uk}} \ and \ {\large
Anca Tureanu$^{(c),}$\footnote{anca.tureanu@helsinki.fi} }

\vspace{5mm} 
$^{(a)}$
{\it
Departamento de F\'{\i}sica, Universidade Estadual de
Londrina\\ Caixa Postal 6001,
Londrina-Paran\'a, Brazil}

\vspace{0.2cm} $^{(b)}$ {\it Department of Mathematics,
Heriot-Watt University\\ Colin Maclaurin Building, Riccarton,
Edinburgh EH14 4AS, UK\\ Maxwell Institute for Mathematical
Sciences, Edinburgh, UK\\ The Higgs Centre for Theoretical Physics, Edinburgh, UK}

\vspace{0.2cm} $^{(c)}$ {\it
Department of Physics, University of Helsinki\\
P.O. Box 64, FI-00014 Helsinki, Finland}

\end{center}

\vspace{0.1in}

\begin{abstract}
We revisit the problem of stability of string vacua involving hyperbolic orbifolds using methods from homotopy theory
and K-homology. We propose a definition of Type~II string theory on such
backgrounds that further carry stratified systems of fiber bundles,
which generalise the more conventional orbifold and symmetric string backgrounds, together with a classification of wrapped branes by a suitable generalized homology theory. For spaces stratified fibered over hyperbolic orbifolds we use
the algebraic K-theory of their fundamental groups and Quinn homology
to derive criteria for brane stability in terms of an Atiyah-Hirzebruch type spectral sequence with its lift to K-homology. Stable D-branes in this setting carry stratified charges which induce new additive structures on the corresponding K-homology groups.
We extend these considerations to backgrounds which support $H$-flux, where we use K-groups of twisted group algebras of the fundamental groups to analyse stability of locally symmetric spaces with K-amenable isometry groups, and derive stability conditions for branes wrapping the fibers of an Eilenberg-MacLane spectrum functor.
\end{abstract}
\vfill



\newpage

\setcounter{footnote}{0}
\setcounter{table}{0}

{\baselineskip=12pt
\tableofcontents
}

\bigskip

\section{Introduction and summary}

Backgrounds of string theory with group actions have a distinctive place in the landscape of string vacua, as they provide a procedure for generating new backgrounds from old ones. For example, just as strings can propagate
on spaces, they can also propagate on orbifolds~\cite{Dixon,Sharpe,Pantev}
which lead to many novel stringy effects, particularly in the
description of D-branes. Moreover, all supergravity backgrounds which
preserve more than half the supersymmetry are
homogeneous~\cite{FigueroaO'Farrill}, and symmetric spaces constitute
a special class wherein the classification of these backgrounds can be
substantially progressed.

In this paper we revisit the problem of supersymmetry of string theory backgrounds involving hyperbolic orbifolds; in fact we analyse the weaker problem of stability of these backgrounds. In Sect.~\ref{Hyper} we review the significance of these string vacua and problems associated with their supersymmetry. Starting from features of string propagation in orbifolds and in symmetric spaces, we then consider modifications of these backgrounds wherein they are regarded as the bases of stratified systems of fiber bundles, and more specifically of spaces stratified fibered over hyperbolic orbifolds. We propose a definition of Type~II string theory on these spaces and analyse their stability in terms of wrapped branes, which constitute novel extensions of string theory on orbifold and symmetric backgrounds. The existence of a stratification on a background $\cB$ is of course neither novel nor exotic: If $\cB$ is endowed with the structure of a simplicial complex, then the strata are just the open simplices. However, the more pertinent example to bear in mind throughout this paper is that of an orbifold, which is a disjoint union of strata with each stratum a connected component of points with the same isotropy group (up to conjugacy).

In this paper we shall
introduce and emphasise new methods from homotopy theory for the
classification of branes on these broader classes of singular backgrounds,
which are techniques that have received considerably less attention in the physics literature compared to e.g.~K-theory. The basic idea is that by considering these backgrounds and their branes in stratified families we can probe their stability by endowing them with ``local coefficient systems'', and then classify them by a suitable homology theory on the category of spaces with stratified systems of fiber bundles. This is homology with \emph{twisted} and stratified spectrum coefficients developed by Quinn~\cite{Quinn}; we give a general overview of Quinn homology in Sect.~\ref{sec:homology} using the language of homology loop spectra, paying particular attention to an Atiyah-Hirzebruch type spectral sequence and its relation to singular or \u{C}ech homology. As we discuss in the following, Quinn homology for stratified systems of fiber bundles plays a role analogous to Bredon homology for orbifolds.

In Sect.~\ref{stability} we discuss the classification of stable D-branes in spaces stratified fibered over hyperbolic orbifolds. We use a suitable Atiyah-Hirzebruch type spectral sequence and stratified systems of abelian groups over hyperbolic manifolds to derive rigorous stability criteria. As the initial homology charges are have coeffcients valued in stratified systems of groups, stable D-branes carry charges which induce new additive structures on K-homology. For backgrounds with strongly virtually negatively curved fundamental groups $\Lambda$, we give an explicit description of the homology groups in which the obstructions to D-brane stability lie in terms of the algebraic K-theory of $\Lambda$. We also find a broad class of novel examples of unstable backgrounds.

Finally, in Sect.~\ref{KK-groups} we extend our considerations to examples involving Type~II D-branes in $H$-flux backgrounds, using their description in terms of KK-groups and the twisted K-homology of continuous trace $C^*$-algebras. In the case of locally symmetric spaces with K-amenable isometry groups, we provide an explicit analysis of stability conditions using K-groups of the twisted group $C^*$-algebra of the fundamental group of the background. In particular, when the locally symmetric background is an Eilenberg-MacLane space, we examine conditions under which it can be extended via a stratified system of abelian groups through a suitable Atiyah-Hirzebruch type spectral sequence as a non-trivial element of the homology groups; stability here is formulated in terms of filtration structured spectra associated to the fibers of the Eilenberg-MacLane spectrum functor.

\section{String orbifolds, stratified fiber bundles and branes}
\label{Hyper}

{\bf Orbifolds. \ }
Let $\Lambda$ be a discrete group acting properly on a smooth manifold
$X$ with finite stabilizers. A Type~II string orbifold is then constructed
by taking a worldsheet $\Sigma$, which is a compact oriented
two-manifold, and a collection of smooth maps
$\phi:\Sigma\to X$. To include twisted sectors into the string theory, we
consider a larger space of fields which is a groupoid; the objects of
the groupoid label the twisted sectors and 
are pairs $(P,\tilde\phi\, )$ consisting of a principal $\Lambda$-bundle
$P\to\Sigma$ with a $\Lambda$-equivariant map
$\tilde\phi:P\to X$, while the morphisms $(P,\tilde\phi\, )\to
(P',\tilde\phi{}'\, )$ are given by isomorphisms $f:P\to P'$ of
principal $\Lambda$-bundles such that $\tilde\phi=\tilde\phi{}'\circ
f$. This
defines the string theory on the global quotient spacetime
$\cB=[X/\Lambda]$, which can be presented as the action groupoid $\Lambda\times X\rightrightarrows X$ whose objects are the points
$x\in X$ and whose morphisms $x\to y$ are the group elements $g\in\Lambda$
for which
$g\cdot x=y$; the groupoid structure keeps track of the isotropy subgroups
$\Lambda_x:=\{h\in\Lambda\, |\, h\cdot x=x\}\subset\Lambda$ for all $x\in X$, and a
string field $\phi:\Sigma\to\cB$ is given by a pair $(P,\tilde\phi\,)$
as above. A geometric realization of this quotient is given by
the Borel construction $|\cB|:=\underline{E}\Lambda\times_\Lambda X$, where
$\underline{E}\Lambda$ is a contractible space with a free $\Lambda$-action; if $X$ is contractible then the orbifold fundamental group is $\pi_1(\cB)=\pi_1(|\cB|)=\Lambda$. For
further details and background on orbifolds, see e.g.~\cite{Adem}.

We can also admit as spacetimes more general orbifolds $\cB$ which are
not global quotients by discrete groups. Generally, orbifolds are
presented as groupoids, so that points can have automorphisms $\Lambda_x$ as above, or alternatively as smooth real Deligne-Mumford
stacks; see~\cite{Heinloth} for an introductory exposition of results
that we shall use throughout this paper. Recall that a stack is an object in the 2-category of sheaves
of groupoids on the category of smooth manifolds with respect
to the usual Grothendieck topology given by open coverings. Any
manifold $X$ is itself a stack which associates to each test manifold
$T$ the (trivial) groupoid of smooth maps $X(T)=C^\infty(T, X)$. The global quotient stack
$[X/\Lambda]$ associates to $T$ the groupoid $[X/\Lambda](T)$ of pairs $(P\to
T,\tilde\phi\, )$ as above; if $\Lambda$ acts freely with
quotient manifold $X/\Lambda$, then there is a natural isomorphism of
stacks $[X/\Lambda]\cong X/\Lambda$ or alternatively a
Morita equivalence between the action groupoid $\Lambda\times X\rightrightarrows X$ and the unit groupoid $X/\Lambda\rightrightarrows X/\Lambda$. The maps between
two stacks $\cB\to\cB'$ form the groupoid $\Homsf(\cB,\cB'\,)$ of maps between sheaves of groupoids;
the objects of $\Homsf(\cB,\cB'\,)$ are morphisms $f:\cB\to \cB'$ of stacks and the
morphisms of $\Homsf(\cB,\cB'\,)$ are 2-morphisms $f_1\Rightarrow f_2$
between morphisms $f_1,f_2 :\cB\to \cB'$. The Yoneda embedding then gives a Morita
equivalence of groupoids $\cB(T)\cong\Homsf(T,\cB)$. The string fields can thus be
regarded as maps $\phi:\Sigma\to\cB$ of stacks, and the groupoid of all
such maps includes twisted sectors of the orbifold string theory. If
the stack
$\cB$ is presented as a groupoid $\cG=(\cG_1\rightrightarrows\cG_0)$, then the collection of $n$-tuples of
composable morphisms $\cG_n=(\cG_1)^{\times_{\cG_0}\, n}$ is a simplicial space with geometric
realisation $|\cB|=\coprod_n\, \cG_n\times\Delta^n/\sim\,$, where
$\Delta$ denotes the standard simplicial 2-category of ordered simplices and the
equivalence relation $\sim$ identifies the face and degeneracy maps of $\cG$ and $\Delta$; throughout
$\Delta^n$ denotes the standard $n$-simplex. The set of
isomorphism classes of objects of the groupoid is called the orbit space of the
orbifold, or the coarse moduli space of the stack. For further details concerning stacks in relation to string orbifolds, see~\cite{Sharpe}.

Other geometric data, such as metrics, connections and supergravity form fields, can also be defined
in this setting. In particular, D-branes wrapping $\cB$ carry gauge
fields and require a notion of Chan-Paton bundles on stacks. A vector
bundle on a stack $\cB$ is a map of stacks from $\cB$ to the stack of
vector bundles whose evaluation on a test manifold $T$ is the groupoid
of vector bundles over $T$ and isomorphisms. Alternatively, given a
groupoid presentation $\sfs,\sft:\cG_1\rightrightarrows\cG_0$ for
$\cB$, a vector bundle on $\cB$ is a vector bundle $\cE\to \cG_0$
together with an isomorphism $\sfs^*\cE\xrightarrow{\sim}\sft^*\cE$ of
bundles over $\cG_1$ which satisfies a cocycle condition over $\cG_2=
\cG_1\times_{\cG_0}\cG_1$. Finite rank vector bundles on a groupoid and
their morphisms form a symmetric monoidal category; the Grothendieck group of this category computes the K-theory
of the stack $\cB$.

\bigskip

{\bf Stratified fiber bundles. \ }
In this paper we will be concerned with a homotopy theoretic
generalization of the class of string backgrounds given by
orbifolds. For this, recall that a stratification of a space $\cB$ is
a locally finite decomposition $\cB=\coprod_i\, \cB_i$ into pairwise
disjoint and locally closed submanifolds $\cB_i$, called strata. The
usual homotopy lifting property which defines a fiber bundle is generalized to the stratified setting
as follows: A map $p:X\to\cB$
between stratified spaces is called a \emph{stratified fiber bundle} if, for
any map $f:E\to X$, every stratum preserving homotopy
$F:E\times\Delta^1\to \cB$ from $F|_{0}= p\circ f$ lifts to a stratified homotopy
$\tilde F:E\times\Delta^1\to X$ from $\tilde F|_0= f$ with $p\circ\tilde
F=F$. See e.g.~\cite{Hughes} for further details and properties.

The global quotient orbifolds describing string backgrounds fit into this framework
in the following way. Suppose that $\Lambda$ is a finite group acting
smoothly on a manifold $X$. For a subgroup $H\subseteq \Lambda$, let
$X_{(H)}$ be the $\Lambda$-invariant submanifold of $X$ consisting of
all points $x\in X$ whose isotropy subgroup $\Lambda_x$ is conjugate
to $H$; then $X_{(\Lambda)}$ is the fixed point locus of the
$\Lambda$-action, while $X_{(1)}$ is the set of points on which
$\Lambda$ acts freely, i.e., the regular points which form a dense open subset of $\cB=[X/\Lambda]$. The distinct sets $X_{(H)}$ depend only on the
conjugacy class of $H\subseteq\Lambda$ and give the strata of
the stabilizer stratification of $X$; the corresponding quotients
$X_{(H)}/\Lambda$ give the orbit stratification of the orbit space
$X/\Lambda$ with the factor topology. Then the orbit map $X\to X/\Lambda$ is a stratified
fiber bundle of coarse moduli spaces underlying the map of stacks
$X\to\cB$ which defines a principal $\Lambda$-bundle over $\cB$. A similar construction holds for stacks which are
presentable as local quotient groupoids, i.e., as groupoids which admit a
countable open cover $\{\cU_i\}$ such that the restriction to each
$\cU_i$ is Morita equivalent to an action groupoid $\Lambda_i\times U_i\rightrightarrows U_i$~\cite{Freed}.

\bigskip

{\bf Symmetric spaces. \ }
An important generalization of this last construction of stratified
fiber bundles is to the case where $\Lambda$ is replaced by a compact Lie
group $\Gamma$. Then one can regard the covering projections $X\to
X/\Gamma$ as a collection of fiber bundles by the
property that each orbit map $X_{(H)}\to X_{(H)}/\Gamma$, for $H\subseteq \Gamma$, is a smooth fiber
bundle with fibers $\Gamma/H$.

This extension pertains in particular to the class of homogeneous Type~II string
backgrounds given by (locally) symmetric spaces. For this, let $G$ be a connected
semisimple Lie group without compact factors. Let $\Gamma\subset G$ be a
maximal compact subgroup, and let $\Lambda\subset G$ be a cocompact
torsion-free lattice. Then the contractible manifold $X=G/\Gamma$ carries a $G$-invariant
Riemannian metric and is a non-compact symmetric space. The
discrete group $\Lambda$ acts freely and properly on $X$ via $\lambda\cdot(g\,\Gamma):= (\lambda\, g)\, \Gamma$ for $\lambda\in\Lambda$ and $g\in G$, and the
closed manifold $\cB=\Lambda\backslash X$ is a compact locally symmetric
space with the factor topology which is a model for the classifying stack $\underline{B}\Lambda=\underline{E}\Lambda/\Lambda \cong
[{\rm pt}/\Lambda]$ of principal $\Lambda$-bundles with $X=\underline{E}\Lambda$ a model for the universal space for proper $\Lambda$-actions. If $\Lambda$ has torsion, then the
$\Lambda$-action on $X$ need not be free since the isotropy subgroups
are $\Lambda_x=(x^{-1} \, \Lambda \, x)\cap\Gamma$  for $x\in X$ (which are
finite as they are discrete subgroups of the compact group $\Gamma$);
in this instance $\cB$ is
a stratified space as above. Such backgrounds preserve more than half of the supersymmetry; backgrounds
preserving a substantial amount of supersymmetry play a crucial role
in string theory, particularly in gauge theory/gravity
correspondences.

\bigskip

{\bf Stability of hyperbolic orbifolds. \ }
In the case that $G$ is a connected Lie group of real rank one and $\cB= \Lambda\backslash X=\Lambda\backslash G/\Gamma$ is a compact locally symmetric Riemannian manifold of dimension $n$ with negative sectional curvature, the covering manifold $X$ is a real hyperbolic space $H^n$. We shall mostly focus on these spaces as they yield a class of backgrounds $\cB$ which are potentially compatible with supersymmetry in our constructions to follow; in particular, as supergravity solutions they naturally arise as near-horizon geometries of branes.

The $n$-dimensional real hyperbolic space can be represented as the symmetric space $H^n=G/\Gamma$, where $G=SO_1(n,1)$ 
and $\Gamma =SO(n)$ is a maximal compact subgroup of $G$. They admit
Killing spinors \cite{BF,FY,LP,Lu}, but they have infinite volume with
respect to the Poincar\'e metric and hence do not support field
configurations with normalizable zero modes. On
the other hand, consistent field theories can be obtained by taking
backgrounds $\cB$ with the topology of coset spaces
$H^n/\tilde\Lambda$, where $\tilde\Lambda$ is a discrete subgroup of
$G$ acting isometrically on $H^n$. The question of whether such
quotient spaces admit Killing spinors and preserve supersymmetry is
addressed for some examples of finite volume hyperbolic spaces
in~\cite{Russo}. Moreover, a version of the gauge theory/gravity
correspondence states that string theory (or M-theory) compactified on spaces of
the form $AdS_{d+1}\times ({H}^n/\tilde\Lambda) \times {S}^{k}$ defines a $d$-dimensional conformal field theory with $SO(k+1)$ global 
symmetry. We mention three such particular cases of supergravity
solutions from~\cite{Russo} which will be pertinent later on. Firstly, the 11-dimensional
supergravity solution of the form $AdS_5\times (H^2/\tilde\Lambda) \times {S}^4$ is dual to a $d=4$ non-supersymmetric conformal
field theory with $SO(5)$ global symmetry group; it carries M5-brane
charge and is expected to be related to the six-dimensional $(2,0)$ or
$(1,0)$ superconformal field theories. Secondly, the 11-dimensional 
supergravity solution $AdS_4\times (H^3/\tilde\Lambda) \times {S}^4$
has M5-brane charge, and thus its dual field theory is a $d=3$ conformal field theory 
associated with the six-dimensional $(2,0)$ superconformal theory with
an internal global symmetry group $SO(5)$. Thirdly, the Type~IIB supergravity
solution $AdS_3\times (H^2/\tilde\Lambda) \times {S}^5$ has D3-brane charge and it should be dual to some non-supersymmetric $d =2$ conformal field theory related to $\mathcal{N}=4$ supersymmetric
Yang-Mills theory in four dimensions; these models fall into the class
of regular string compactifications (modulo possible orbifold points)
which have small $\alpha'$ string corrections and are holographically
dual to $\mathcal{N}=0$ conformal field theories in four dimensions~\cite{Russo,kach,NV}.

Various string backgrounds can be presented by means of direct products of spaces containing 
real hyperbolic space 
forms as factors; for example the space forms 
$AdS_{3}\times H^{2}\times {H}^2\times {S}^4$ and 
$AdS_{2}\times {H}^{2}\times{H}^3\times{S}^4$ appear as solutions of 11-dimensional supergravity. However, it is easy to show that any connected locally Riemannian product $M=H^k\times H^l$ cannot leave any unbroken supersymmetry. For this, suppose that $M$ admits a Killing spinor $\psi\neq0$ with Killing number $\mu \neq 0$; then $M$ is
locally irreducible~\cite{Friedrich}. Let
${\xi}$ and $ {\eta}$ be (pullbacks to $M$ of) vector fields on $H^k$ and
$H^{l}$ respectively; then $\xi\cdot\eta=0$ and $\eta\cdot\xi=0$. The Killing spinor equations at each point of $M$ then read as 
$
\nabla_{\xi}\psi = \mu \, {\xi} \cdot \psi$ and similarly for $\nabla_\eta$, where
$4\mu^2 =  \big(n\, (n-1)\big)^{-1}\, R
$
and $R$ is the scalar curvature (here $n=k+l$). Thus we have
\begin{equation*}
R(\xi,\eta)\, \psi = \big(\nabla_{\xi}\, \nabla_{\eta} -
\nabla_{\eta}\, \nabla_{\xi} - \nabla_{[{\xi}, {\eta}]} \big)\psi
= \mu^2\, [{\eta},{\xi}]\cdot \psi = 0 \ ,
\end{equation*}
and since $R \neq 0$ it follows that $\psi =0$, which is a
contradiction. Other compactifications can be obtained by replacing
hyperbolic space forms $H^n$ by any Einstein space $M$ of the same
negative curvature and with Killing spinors; in fact, if $M$ is a connected Riemannian $Spin$
manifold admitting a non-trivial Killing spinor with non-zero Killing
number, then $M$ is an Einstein space~\cite{Friedrich}. In particular,
one can consider $(H^k\times H^l)/\tilde\Lambda$ and ask if the
orbifold by the discrete group $\tilde\Lambda$ restores
supersymmetry. But even if non-supersymmetric, these solutions are
regular everywhere (being a direct product of Einstein spaces) and
$\alpha'$ corrections can be made small for sufficiently large radius
of the compact parts. Hence like the Type~IIB supergravity backgrounds
$AdS_3\times
{H}^2\times {S}^5$ and $AdS_2\times {H}^3\times{S}^5$ of~\cite{Russo}
with D3-brane charge and constant dilaton, these spaces could still
constitute consistent backgrounds for string compactification, and in
particular for the construction of conformal field theories. In
general a definitive statement about the stability of supergravity
solutions requires study of the spectral properties of certain operators,
which is largely undeveloped as of yet.

\bigskip

{\bf Spaces stratified fibered over hyperbolic orbifolds. \ }
To address these stability issues, we add more structure to this
construction: to detect possible instabilities we consider the background $\cB$ in suitable families which ``probe'' the structure of the orbifold. Following~\cite[Sect. 8]{Quinn}, let
$\cB=\cB^n\supseteq\cB^{n-1}\supseteq \cdots\supseteq \cB^0$ be a
closed finite filtration of the space $\cB$; then one obtains a
stratification of $\cB$ by setting $\cB_i:=\cB^i-\cB^{i-1}$ for
$i=0,1,\ldots,n$, where $\cB^{-1}:=\varnothing$. We will usually use a
skeletal filtration whereby $\cB^i$ is the $i$-skeleton of a CW-decomposition of $\cB$; then the strata $\cB_i$ are the open $i$-simplices. A map
$p:X\to\cB$ is called a \emph{stratified system of fiber bundles} on $\cB$ if
the restriction to each stratum $p|:p^{-1}(\cB_i)\to\cB_i$ is an ordinary
fiber bundle, and each term in the filtration $\cB^i$ is a $p$-neighborhood
deformation retract subset of $\cB$, i.e., there exists a neighborhood
of $\cB^i$ in $\cB$, with a continuous deformation into $\cB$ that
preserves all strata and is covered by a deformation of its fibers in $X$. In the case
of string orbifolds, with the discrete group $\Lambda$ acting
cellularly on $X$, the projection of the geometric realisation $|\cB| =\underline{E}\Lambda\times_\Lambda
X$ to the second factor defines a stratified system of fiber bundles
$p:|\cB| \to X/\Lambda$~\cite{Talbert}; in this case there is a
homotopy equivalence $p^{-1}(\sigma)\cong \underline{B}\Lambda_\sigma$ for any simplex
$\sigma$ of $\cB=[X/\Lambda]$.

Now let $M$ be a closed connected Riemannian manifold with strictly negative
sectional curvatures and let $\{\cX^j\}_{j =1}^\infty$,\, $ \cX^1
\subset \cX^2 \subset \cdots $ be a sequence of connected
compact smooth manifolds. Let $\tilde\Lambda$ be a finite group acting
on $M$ freely via isometries and on each $\cX^{j}$ via smooth
maps; the action of $\tilde\Lambda$ on $\cX^j$ need not be free. Assume moreover that the smooth embedding
$\cX^{j}\subset \cX^{j+1}$ is both $\tilde\Lambda$-equivariant and
$j$-connected, i.e., its homotopy groups vanish in degree $\leq j$. Let $\cX^{\infty} = \bigcup_{j=1}^{\infty}\, \cX^{j}$ with the direct limit topology. The induced action of $\tilde\Lambda$ on
$\cX^\infty$ is free. Let $X^j$ be the orbit
space of $\cX^{j}\times M$ under the diagonal action of
$\tilde\Lambda$, and set
\begin{equation*}
X^\infty = \lim_{\stackrel{\scriptstyle\longrightarrow}
{\scriptstyle j}}\, X^j\ .
\end{equation*}
Let $\cB$ be the orbit space $M/\tilde\Lambda$ and
$p_j :  {X}^{j}\rightarrow {\cB}$ the map
induced from the canonical projection of $
\cX^{j}\times M$ onto $M$. The set $\{p_{j}\}_{j=1}^\infty$ is a
collection of stratified
systems of fiber bundles on $\cB$ in the sense described above. 
For our ensuing applications we will have to define a homotopy class of maps
$p_{\infty}:X^\infty\to\cB$; in the following we
write $X,p$ for $X^\infty,p_\infty$ in order to simplify notation.

\bigskip

{\bf Stratified string fields and branes. \ }
In Type II string theory on the hyperbolic orbifold background $\cB$, one
needs to endow $\cB$ with extra geometric data, such as a
$Spin$ structure; the manifold $\cB$ admits $Spin$ structures if and only if its first and second Stiefel-Whitney classes vanish. In what follows we shall analyse to what extent this is compatible with stability and the
stratification structure of $p:X\to\cB$. In the worldsheet theory this
is accomplished as follows.
\begin{definition}\label{def:stratstring}
Let $p:X\to\cB$ be a collection of stratified fiber bundles over a
hyperbolic orbifold. A \emph{Type II stratified string orbifold} is a quadruple
$(\Sigma,\phi;\widehat{\Sigma} ,\widehat{\phi}\, )$ consisting of a string field
$\phi:\Sigma\to\cB$, a stratified system of fiber bundles
$\widehat{p}:\widehat{\Sigma}\to\Sigma$ endowed with a $Spin$ structure, and a stratum preserving lift of $\phi$ to an equivariant map $\widehat{\phi}: \widehat{\Sigma}\to X$ of the same homotopy type such that the diagram of maps of stacks $\widehat{\Sigma}\to\cB$
\begin{equation}
\xymatrix@=15mm{
\widehat{\Sigma} \ \ar[d]_{\widehat{p}} \ar[r]^{\widehat{\phi}} & \ X \ar[d]^p \\
\Sigma \ \ar[r]_\phi & \ \cB
}
\label{eq:stratstringdiag}\end{equation}
commutes.
\end{definition}
Let us unpack Def.~\ref{def:stratstring}. As before, the string
field $\phi:\Sigma\to\cB$ is an oriented principal
$\tilde\Lambda$-bundle $P\to\Sigma$ and a $\tilde\Lambda$-equivariant
map $P\to M$, and similarly $\widehat{\phi}:\widehat{\Sigma}\to X$ is a principal
$\tilde\Lambda$-bundle $\widehat{P} \to\widehat{\Sigma}$ and a
$\tilde\Lambda$-equivariant map $\widehat{P} \to\cX^\infty\times M$; the
pullback by $\widehat{p}:\widehat{\Sigma}\to\Sigma$ induces a morphism of principal
$\tilde\Lambda$-bundles $\widehat{p}{\,}^*:P\to \widehat{P}$. If the stratification of the
worldsheet is $\Sigma=\coprod_i\,
\Sigma_i$, then commutativity of the diagram
(\ref{eq:stratstringdiag}) implies the stratum preserving property
$\phi(\Sigma_i)=(p\circ\widehat{\phi}\, )(\widehat{p}\,^{-1}(\Sigma_i))\subseteq
\cB_i$. The collection of all quadruples
$(\Sigma,\phi;\widehat{\Sigma} ,\widehat{\phi}\, )$ forms a groupoid in
the obvious way; they provide the worldsheet ``probes'' of the structure of the background~$\cB$.

In the particular instance
where $M$ involves hyperbolic space forms $H^n$, supersymmetry (or more
generally stability) of the
background $\cB$ and the brane charges it supports can be analysed as discussed above. 
In Sect.~\ref{stability} we shall develop
stability conditions for branes in these backgrounds. Our new
impetus will be a modification of the usual spectral sequence
extension problem by regarding $\cB$ as the base of a stratified
fiber bundle, which incorporates the stratifications $\widehat{\phi}$ of open string
fields from Def.~\ref{def:stratstring} when the worldsheet $\Sigma$
has a boundary $\partial\Sigma\neq\varnothing$; the existence of the
equivariant lift $\widehat{\phi}$ already constrains the homology of
$\Sigma$. For this, we shall propose that branes in these
backgrounds are classified by a suitable homology theory on the
category of spaces with stratified systems of fiber bundles (in the
sense of the Eilenberg-Steenrod axioms for homology). This is the
spectral sheaf homology of Sect.~\ref{sec:homology} below which defines a
functor of simplicial maps $p:X\to\cB$; we think of $p$ as defining a
``local coefficient system'' over $\cB$ and define the associated
twisted homology, which has a concrete realization in terms of chain
complexes. Amongst other features, we shall see that the resulting stability conditions play
an important role in assessing the stability of the hyperbolic orbifold
backgrounds discussed above.

\section{Quinn homology}
\label{sec:homology}

{\bf Homology loop spectra. \ }
We shall explain in some generality the homotopy theoretic approach
to calculating the pertinent
homology groups which arise in the following; see~\cite{Bousfield} for
further definitions and discussion. In this paper we shall always
assume that the background $\cB$ is a simply connected finite-dimensional
CW-complex. For
the present discussion we assume for simplicity that 
$\cB$ is compact, otherwise it should be replaced by its one-point
compactification $\cB_+$ in all considerations below. Let $\cB_0$ be a
subspace of $\cB$. Let
$\scrG=\{\scrG_i\}_{i\in\IZ}$ be an arbitrary loop spectrum with
canonical homotopy equivalences of based CW-spaces
$\scrG_i\to\Omega\scrG_{i+1}$, where $\Omega$ denotes the based loop
space. 

Then the (generalized) homology groups for the finite CW-pair $(\cB,\cB_0)$ with coefficients
in $\scrG$ are defined for $k\in\IZ$ by~\cite{Whitehead}
\begin{eqnarray}
H_k(\cB,\cB_0; \scrG)  =  \lim_{\stackrel{\scriptstyle\longrightarrow}
{\scriptstyle j}}\, \pi_{j+k}((\cB/\cB_0) \wedge\scrG_j) \ ,
\label{OM2}
\end{eqnarray}
where the factor space $X \wedge Y:= X \times Y / X
\vee Y$ is the smash product of two spaces $X$ and $Y$ with
base points $x_0$ and $y_0$ which is obtained
from the product $X\times Y$ by collapsing the subspace $(X\times
\{y_0\})\cup (\{x_0\}\times Y)$ to a single point. For
the directed limit (\ref{OM2}) (which is a categorical colimit in the category of spectra) one uses the map
\begin{equation*}
\pi_{j+k} ((\cB/\cB_0) \wedge \scrG_j) \ \longrightarrow \ 
\pi_{j+k+1}((\cB/\cB_0) \wedge \scrG_{j+1})
\end{equation*}
induced by the mapping $\Sigma \scrG_j \rightarrow \scrG_{j+1}$
as the composition
\begin{equation}
\xymatrix{
\pi_{j+k} ((\cB/\cB_0) \wedge \scrG_j ) \ 
\ar[r]^{\!\!\!\!\!\!\!\!\!\!\!\!\!\Sigma}
& \ \pi_{j+k+1}(\Sigma((\cB/\cB_0) \wedge \scrG_{j+ 1})) \ar@{=}[d]
& {}
\\
& \pi_{j+k+1}((\cB/\cB_0) \wedge\Sigma \scrG_{j} ) \ \ar[r]
& \ \pi_{j+k+1}((\cB/\cB_0) \wedge \scrG_{j+ 1} )
}
\label{AH}
\end{equation}
where for a space $X$ the suspension $\Sigma X$ is the space obtained
from the product $X\times \Delta^1$ by collapsing the subspaces
$X\times\{0\}$ and $X\times\{1\}$ to single points, and we have used
the continuous structure maps $\scrG_j\to\Omega\scrG_{j+1}$. 

One can alternatively define the homology groups $H_\sharp(\cB,\cB_0;
{\scrG})$ directly as the homotopy groups of a loop spectrum
$(\cB/\cB_0)\otimes\scrG$ by setting
\begin{equation*}
((\cB/\cB_0) \otimes {\scrG})_ i = \lim_{\stackrel{\scriptstyle
    \longrightarrow}{\scriptstyle j}} \, \Omega^j((\cB/\cB_0) \wedge
  {\scrG}_{i+j}) \ ,
\end{equation*}
where the maps in the directed system are induced from the structure
maps ${\scrG}_i\rightarrow { \Omega}{\scrG}_{i+1}$ of the loop spectrum
$\scrG$.

In general the relation between homotopy and homology functors is very subtle,
even though there are natural homomorphisms $\pi_j(\cB)\to H_j(\cB)$;
for example, for $j\geq1$ there is the Hurewicz homomorphism
$\pi_j(\cB)\to H_j(\cB;\IZ)$. While the homology groups $H_j(\cB)$ are
relatively computable, the homotopy groups $\pi_j(\cB)$ are typically
not as there is no analogue of the Mayer-Vietoris sequence that
is implied by the Eilenberg-Steenrod axioms for homology
theories. This means that the homology groups defined in this way are in general
difficult to compute explicitly.

\bigskip

{\bf Spectral sheaves. \ } 
Following~\cite{James} let us consider a fiber bundle over $\cB$ with section
\begin{equation*}
\xymatrix@C=20mm{
 \scrG \ \ar@/^/[r]^{\pi} &
  \ \cB \ar@/^/[l]^s
}
\end{equation*}
so that $\pi \circ s={\rm id}_\cB$. The loop space
$\Omega_\cB(\scrG)$ over $\cB$ is the space of maps
$F:\cB\times\Delta^1\to\scrG$ which commute with the projection $\pi$ to
$\cB$, and which coincide with the section $s$ when restricted to
$\cB\times\{0\}$ and $\cB\times\{1\}$. Then $\Omega_\cB(\scrG)$ is
again a fiber bundle over $\cB$ with section and the fibers of
$\Omega_\cB(\scrG)\to \cB$ are loop spaces of the fibers of $\scrG
\to\cB$. A \emph{spectral sheaf} is a sequence of fiber bundles with
section $\{\scrG_j\}_{j\in\IZ}$ and morphisms $\alpha_j:\scrG_j \to \Omega_\cB(\scrG_{j+1})$. For a subspace $\cB_0\subset\cB$, these structure maps define maps~\cite{Quinn}
\begin{equation*}
\scrG_j\,\big/\, \iota(\cB) \cup\pi^{-1}(\cB_0) \ \xrightarrow{ \ \alpha_j \ } \ \Omega_{\cB}\big( \scrG_{j+1}\big) \,\big/\, s(\cB)\cup\pi^{-1}(\cB_0) \ \longrightarrow \ \Omega\big(\scrG_{j+1}\,\big/\, s(\cB)\cup\pi^{-1}(\cB_0)\big) \ ,
\end{equation*}
where the second map is the inclusion of based loop spaces and
$s(\cB)\cup\pi^{-1}(\cB_0)$ is taken to the base point. Then there is a homology loop spectrum defined by
\begin{equation*}
\scrQ_k(\cB,\cB_0;\scrG) = \lim_{\stackrel{\scriptstyle\longrightarrow}{\scriptstyle j}}\, \Omega^j\big( \scrG_{k+j}\,\big/\, s(\cB)\cup\pi^{-1}(\cB_0) \big) \ ,
\end{equation*}
and the \emph{Quinn homology groups} $Q_\sharp(\cB,\cB_0; {\scrG})$ are the homotopy groups of this spectrum~\cite{Quinn}; the structure maps on $\scrQ_k(\cB,\cB_0;\scrG)$ are
induced by those for $\scrG$.

\bigskip

{\bf Spectral sequences. \ }  Let us now consider the
Atiyah-Hirzebruch type spectral sequence with constant coefficients
which
is derived in \cite{Whitehead2}. For this, we note that any spectrum $\scrG$ has a natural $j$-connected {cover}
${\scrG}^{(j)} \rightarrow {\scrG}$.
Applying this construction to the fibers of a spectral sheaf
${\scrG} \rightarrow \cB$ gives a sequence of spectral sheaves
\begin{equation}
\cdots \ \longleftarrow \ {\scrG}^{(j)} \ \longleftarrow \
{\scrG}^{(j+1)} \ \longleftarrow \ \cdots
\label{ex}
\end{equation}
which we can regard as a filtration of ${\scrG}$.
Applying the homotopy functor $\pi_\sharp$ to the filtration~{\rm
  (\ref{ex})} gives a spectral sequence which converges to the
spectral sheaf homology $Q_\sharp(\cB; {\scrG})$. 
The $E^2$ terms are given by
$$
E_{i,j}^2=H_i(\cB;\pi_j\scrG)
$$
because the fiber over $\cB$ of ${\scrG}^{(j+1)}\rightarrow {\scrG}^{(j)}$ is the spectral sheaf
${\Omega}_{\cB}^{-j-2}\big(\,\underline{K}(\pi_{j+1}{\scrG},\sharp)\big)$; here
$\underline{K}(-,\sharp)$ is the fiberwise Eilenberg-MacLane spectrum functor from
abelian groups to loop spectra which corepresents the homology
functor, i.e., for any abelian group $\Lambda$ one has
$H_i(\cB;\Lambda)= \pi_i(\cB\wedge\scrH)$ for the spectrum
$\scrH=\{\scrH_i\}_{i\in\IZ}$ defined by taking
$\scrH_i=\underline{K}(\Lambda,i)$. This functor will play an important role again in Sect.~\ref{KK-groups}.

In our specific applications to follow all spectra considered are those of \u{C}ech-type theories, and thus there are isomorphisms between homomorphisms in these theories. The homologies of these spectra with coefficients in sheaves are isomorphic in a unique way, and in particular all constructions for \u{C}ech homology are in one-to-one correspondence with those of singular homology. Hence the differentials of the spectral sequences can be gleamed from those of singular homology.

\section{Branes in spaces stratified fibered over orbifolds}
\label{stability}

{\bf Homology with stratified coefficients. \ } 
Suppose that the background $\cB$ is filtered by closed subsets
$\cB^i\subset \cB^{i+1}$, $i\geq1$. A \emph{stratified system of groups}
  $A$ over $\cB$ consists of neighborhoods $U_i$ of the strata $\cB_i= \cB^i-\cB^{i-1}$,
locally constant systems of groups $A_i:=A(U_i)$ over $U_i$, and for each $i> j$ a homomorphism $\theta_{ij}: A_i\rightarrow A_j$ over $U_i\cap U_j$
such that if $i>j>k$ then $\theta_{jk}\circ\theta_{ij} = \theta_{ik}$ over $U_i\cap U_j\cap U_k$. 

Let $\scrG$ be a
spectrum-valued homotopy invariant functor on the category of spaces, and let $\scrG(Y)$ be the stable
topological pseudo-isotopy loop spectrum associated to a
topological space $Y$; a covariant functor $\scrG$ from spaces to
spectra is homotopy invariant if a homotopy
equivalence of spaces $X\cong Y$ induces a homotopy
equivalence of spectra ${\scrG}(X)\cong {\scrG}(Y)$~\cite{Quinn}. Then to a stratified
system of fiber bundles $p: X \rightarrow {\mathcal B}$ one can
associate the Quinn homology loop spectrum ${\scrQ}({\mathcal B};
\scrG(p))$ with twisted spectrum coefficients and a map of loop spectra $ {\scrQ}({\mathcal B};
\scrG(p))\rightarrow \scrG(X)$, where $\scrG(p)$ is the spectral sheaf
over $\cB$ obtained by applying $\scrG$ to the fibers of $p$
\cite{Quinn}. One can construct an Atiyah-Hirzebruch type
spectral sequence in the following way \cite{Quinn,Farrell}.
The homotopy groups
$\pi_j\scrG(p^{-1}(b))$ corresponding to the point $b\in\cB$ form
stratified systems of abelian groups over $\cB$,
which we denote by $\pi_j\scrG(p)$. In this case the filtration of $\cB$ gives a (homological) spectral
sequence with
\begin{equation}
E_{i,j}^2 = {H}_i({\mathcal B}; \pi_j\scrG(p)) \label{ss}
\end{equation}
which converges to the spectral sheaf homology groups
$Q_{i+j}(\cB;\scrG(p))= \pi_{i+j}{\scrQ}({\mathcal B};
\scrG(p))$. In the case of string orbifolds, i.e., when the stratified
system of fiber bundles is given by a group action as in
Sect.~\ref{Hyper}, this spectral sequence agrees with that
of~\cite{Davis}. In particular, in this instance the spectral sheaf
homology is isomorphic to Bredon equivariant homology~\cite{Talbert}
which naturally realises stringy orbifold cohomology and
stability conditions for fractional D-branes on
orbifolds~\cite{SzaboValentino}.

Let $A$ be a stratified system of abelian groups over $\cB$. The
homology groups $H_j({\mathcal B}; {A})$ can be
calculated as follows. A compact background $\cB$ can be triangulated by a simplicial complex, which is a finite family of closed subsets $\{{\cT}_j\}_{j =0}^n$
which cover $\cB$ together with a family of homeomorphisms $\varphi_j :
\Delta^j\rightarrow {\cT}_j$ where each
$\Delta^j \subset \IR^n$ is a Euclidean $j$-simplex; we assume that each stratum ${\cT}_j$ is a subcomplex. Then a
stratified system $A$ restricted to each open simplex is a
constant system of coefficients and
\begin{equation*}
H({\mathcal B}; A)=\bigoplus_{j\geq0}\, H_j(\cB; A)
\end{equation*}
is the homology of the finite chain complex
\begin{eqnarray*}
\ C_0({\mathcal B}; A) \ \xleftarrow{ \ \partial_1 \ } \ C_1({\mathcal B}; A) \ \xleftarrow{ \ \partial_2 \ } \ \cdots \
\xleftarrow{ \ \partial_n \ } \ C_n({\mathcal B}; A) \ , \qquad
C_j({\mathcal B}; A)  = \bigoplus_{\sigma \in {\cT}_j}\, A(\sigma) \ ,
\end{eqnarray*}
where the simplicial boundary homomorphisms $\partial_j$ are defined by decomposing a $j$-simplex
$\sigma\in \cT_j$ into $(j+1)!$ simplices
of smaller dimension; the vertices of the new simplices are the
centers of gravity of the faces of the original simplex.

\bigskip

{\bf $\boldsymbol{E^2_{0,j}}$-terms. \ } The $E_{0,j}^2$-terms in the spectral sequence
(\ref{ss}) can be evaluated in the following manner. Let
$\Pi$ be the extension of the fundamental group $\pi_1(M)$
determined by the action of $\tilde\Lambda$ on $M$; the
group $\Pi$ is isomorphic to the factor group
$\pi_1(X)/\pi_1(\cX^\infty)$. Let $\scrF (\Pi)$ be the category
with objects the finite subgroups of $\Pi$; for each
element $\beta \in \Pi$ we can determine a morphism
$B_1\rightarrow B_2$ of objects of $\scrF (\Pi)$ if $\beta
\, B_1\, \beta^{-1}\subseteq
B_2$. Then the $E^2_{0, j}$-terms are isomorphic to the direct limit
\begin{equation}
E^2_{0, j} \cong \lim_{\stackrel{\scriptstyle \longrightarrow} {\scriptstyle B\in
\scrF(\Pi)}} \, \pi_{j}\scrG(\cX^\infty/\frp (B)) \ ,
\label{isomorphism}
\end{equation}
where $\frp: \Pi \rightarrow {\tilde\Lambda}$ is the canonical
projection and $\cX^\infty/\frp (B)$ are the corresponding orbit spaces. This isomorphism can
be proven by using a basic result of E. Cartan \cite{Short}.

\bigskip

{\bf Strongly virtually negatively curved groups. \ } 
Let us choose a spectrum-valued functor of spaces $\scrG$ with homotopy groups $\pi_i\scrG(X)=\widetilde{K}_i(\IZ\Lambda)$, where $\Lambda:=\pi_1(X)$. In the case when $\cX^\infty$
is contractible one has isomorphisms~\cite{Farrell}
\begin{eqnarray*}
K_i({\mathbb Z}{\Lambda})\otimes {\mathbb Q} \cong \bigoplus_{j\geq0}\,
H_{j}(\cB; K_{i-j}({\mathbb Z}\Lambda_b)\otimes {\mathbb Q})\ ,
\end{eqnarray*}
where $\Lambda_b =
\pi_1(p^{-1}(b))$ for $b\in {\mathcal B}$. Similarly, by choosing $\pi_i\scrG(X)= Wh_i(\Lambda)$ one has~\cite{Farrell}
\begin{eqnarray*}
Wh_i({\Lambda})\otimes {\mathbb Q} \cong \bigoplus_{j\geq0}\, H_{j}(\cB;
Wh_{i-j}(\Lambda_b)\otimes {\mathbb Q})\ ,
\end{eqnarray*}
where the Whitehead groups $Wh_i(\Lambda)$ are the homotopy groups of the cofiber of the map $\scrQ(\, \underline{B}\Lambda;\scrK(\IZ\Lambda))\to \scrK(\IZ\Lambda)$ between spectral sheaf homology and algebraic K-theory spectra. Here $Wh_{i-j}(\Lambda_b)\otimes {\mathbb Q}$ and
$K_{i-j}({\mathbb Z}\Lambda_b)\otimes {\mathbb Q}$ form
stratified systems of abelian groups over ${\mathcal B}$. 
We therefore say that a group $\Lambda$ is {strongly virtually negatively
curved} if it is isomorphic to the fundamental group $\pi_1 (X)$ when $\cX^\infty$ is contractible.

For example, any discrete cocompact subgroup $\Lambda$ of
$G$, where $G$ is one of the connected non-compact simple split rank one Lie groups with finite center $O(n, 1)$, $U(n, 1)$, $Sp(n, 1)$, or
${F_4}$, is strongly virtually negatively curved \cite{Farrell}. Let $\Gamma\subset G$ be
a maximal compact subgroup; then $G/\Gamma$ is an irreducible non-compact
symmetric space of rank one which up to local isomorphism can be
represented as one of the quotients $H^n= SO_1(n,1)/SO(n)$, $SU(n,1)/U(n)$, $((Sp(n,1)/Sp(n))\times Sp(1)$, $F_{4(20)}/Spin(9)$ of dimensions $n$, $2n$, $4n$, $16$ respectively. The double coset space $
{{\mathcal B}} := {\Lambda}\backslash G/{\Gamma}$ is a locally symmetric space
 which is in particular a compact Riemannian manifold with fundamental group $\Lambda$. The algebraic
K-theory of $\Lambda$ can thus be calculated in
terms of the stratified homology of $
{{\mathcal B}} $ through the isomorphisms~\cite{Farrell}
\begin{equation}
K_i({\mathbb Z}{\Lambda})\otimes {\mathbb Q} \cong  \bigoplus_{j\geq0}\,
H_j(\cB ;  \cV_{i-j})\ , \label{Hom}
\end{equation}
where $\cV_i$ is a stratified system of $\mathbb Q$-vector spaces
over $\cB$ such that the vector space
$\cV_i({\Lambda} \,g\, {\Gamma})$ corresponding to the double coset ${\Lambda}
\,g\, {\Gamma}$ for $g\in G$ is isomorphic to $K_i({\mathbb Z}({\Lambda} \cap
g\, {\Gamma}\,g^{-1}))\otimes {\mathbb Q}$; as previously ${\Lambda}
\cap (g\,{\Gamma}\,g^{-1})$ is a finite subgroup of $\Lambda$ because $\Gamma$ is compact.

\bigskip

{\bf Stability conditions for D-branes. \ } 
Let us now analyse D-brane
stability in this setting. We introduce D-branes into these backgrounds following the
definitions and conventions of~\cite{Reis}. K-homology cycles encode data that must be
carried by any D-brane, such as a ${\it Spin}^{\mathbb C}$ structure and a complex
vector bundle. A D-brane on a background $\mathcal
B$ of Type~II
string theory is specified by a triple $(\cW,{F},\psi)$, where $\cW$ is a ${\it Spin}^{\mathbb
C}$ manifold regarded as a brane worldvolume, $F$ is the Chan-Paton bundle on $\cW$ with $[F]\in K^0(\cW)$ and $\psi :
\cW \rightarrow {\mathcal B}$ is a continuous map. D-brane charges can
then be
regarded as homology classes of K-cycles $[\cW, F, \psi]\in
K_\sharp(\mathcal B)$. The equivalence relation of vector bundle
modification in K-homology identifies pairs of D-branes for which one
is a spherical fiber bundle over the other, in accordance with our
treatment of the target space $\cB$ as the base of a system of
fiber bundles.

Any free brane (with no lower or higher brane charges) can wrap a
homologically non-trivial cycle in $\cB$; a D-brane $[\cW,F,\psi]$
\emph{wraps} $\cM\subseteq\cB$ if $\psi(\cW)\subseteq\cM$. The
Atiyah-Hirzebruch spectral sequence provides a mathematical algorithm that determines which homology
classes lift non-trivially to K-homology classes, i.e., it determines which
D-branes are unstable and not allowed; in this instance one chooses a spectrum-valued
functor of spaces $\scrG$ such that $\scrG(X)$ is homotopy equivalent to
the
Bott spectrum $\scrK=\{\scrK_i\}_{i\in\IZ}$ defined by taking
$\scrK_i$ to be the infinite unitary group $U(\infty)$ for $i$ odd and
$\IZ\times \underline{B}U(\infty)$ for $i$ even. For other flavors of branes one could
choose the loop spectrum $\scrG(X)$ appropriate to the generalized
homology theory which classifies those branes, although we shall see
that D-branes (in the sense defined above) fit most naturally into the
present framework. The spectral sequence keeps track of the
possible obstructions for a homology cycle to survive to
$E_{i,j}^\infty$, starting from
${H}_i({\mathcal B}; \pi_j\scrG(p))$ in Eq.~(\ref{ss}); the initial
terms are given in Eq.~(\ref{isomorphism}). Given a brane wrapping a cycle in $\cB$ and carrying a stratified system of groups (or local coefficient systems), we can ask if it is stable in the sense that it has a lift to a non-trivial class in the spectral sheaf homology $Q_\sharp(\cB;\scrG(p))$. We analyse this obstruction problem in the case that $p:X\to\cB$ is simplicially stratified.

The first term of the spectral sequence is the relative Quinn homology
\begin{equation}
E_{i,j}^1=Q_{i+j}(\cB^i,\cB^{i-1};\scrG(p)) \ ,
\label{eq:Eij1}\end{equation}
where $\scrG(p)$ is the disjoint union of
$\scrG(p^{-1}({\sigma})) \times\sigma$ over all simplices $\sigma$ of
$\cB$ modulo the equivalence relation which identifies
$\scrG(p^{-1}({\partial_j\sigma}))\times\partial_j\sigma$ with its
image in $\scrG(p^{-1}({\sigma} ))\times\sigma$. Note that the natural projections $\scrG(p^{-1}({\sigma})) \times\sigma\to\sigma$ fit together to give a projection $\pi:\scrG(p)\to\cB$, while the basepoints of the pieces fit together to form sections $s_j:\cB\to\scrG_j(p)$ for each $j\in\IZ$; the structure maps of the spectra $\scrG(p^{-1}({\sigma}))$ also fit together to give structure maps on $\scrG(p)$ that commute with the sections and projections, and hence $\scrG(p)$ is a spectral sheaf over $\cB$. The first differential
\begin{equation*}
\dd^1_{i,j}\, :\, E^1_{i,j} \ \longrightarrow \ E^1_{i-1,j}
\end{equation*}
is the composition of the simplicial boundary homomorphism
$\partial_i:Q_{i+j}(\cB^i,\cB^{i-1};\scrG(p))\to
Q_{i+j-1}(\cB^{i-1};\scrG(p))$ with the pushforward $\kappa_*:
Q_{i+j-1}(\cB^{i-1};\scrG(p))\to
Q_{i+j-1}(\cB^{i-1},\cB^{i-2};\scrG(p))$ induced by the inclusion
$\kappa:(\cB^{i-1},\varnothing)\hookrightarrow(\cB^{i-1},\cB^{i-2})$. The
group $E_{i,j}^1$ parameterizes spectral sheaf homology classes on the
$i$-skeleton of $\cB$ which are trivial on the $i-1$-skeleton; it
classifies branes wrapping $i$-cycles of the stratified system of
fiber bundles $p:X\to\cB$ which carry no lower or higher degree brane
charges. Alternatively, using excision we may write the group (\ref{eq:Eij1}) as the reduced homology
$$
E_{i,j}^1=\widetilde{Q}_{i+j}(\cB_i;\scrG(p)):={Q}_{i+j}(\cB_i, {\rm pt} ;\scrG(p))
$$
which classifies branes on the $i$-th stratum $\cB_i$ of $p:X\to\cB$
which carry no lower charges; such branes further support ordinary
fiber bundles $p|:p^{-1}(\cB_i)\to\cB_i$ in addition to their usual
Chan-Paton vector bundle which generalizes the roles of image branes of ``fractional
branes'' on the covering space of an orbifold. For the simplicial stratification, $\cB_i$ is a
disjoint union of boundaries $\partial_i\sigma$ of simplicial cells
$\sigma\in\cT_i$, each of whose reduced homology can be computed using excision as
the reduced homology of a sphere
$\widetilde{Q}_{i+j}(S^i;\scrG(p))\cong Q_j({\rm pt};\scrG(p))=
\pi_j\scrG(p)$~\cite{Davis}. It then follows from the
Eilenberg-Steenrod additivity axiom that the homology group (\ref{eq:Eij1}) can be identified with the group of singular $i$-chains
$$
E_{i,j}^1= C_i(\cB;\pi_j\scrG(p)) \ .
$$

The second term of the spectral sequence is the homology of the differential $\dd^1$ and is given by Eq.~(\ref{ss}); by our previous results we can compute this term from the spectral sheaf homology groups $Q_i(\cB;\scrK(\IZ\Lambda_b))$ corresponding to algebraic K-theory, where up to conjugation each $\Lambda_b$ forms a sheaf of groups over $\cB$. On the $r$-th term $E_{i,j}^r$, the differential $\dd^r_{i,j}$ has bidegree $(-r,r-1)$ and $E_{i,j}^{r+1}$ is the corresponding homology group. The $E^\infty$ term is the inductive limit
$$ 
E^\infty_{i,j} =
\lim_{{\stackrel{\scriptstyle \longrightarrow} {\scriptstyle r}}}\, E_{i,j}^r \ .
$$ 
If $n=\dim(\cB)$, then $E_{i,j}^r=E_{i,j}^\infty$ for all $r>n$ and
convergence of the spectral sequence means that there exists an
ascending filtration $F_{m,m-i}Q_m(\cB;\scrG(p))$ of
$Q_m(\cB;\scrG(p))$, with $0\leq m\leq n$, such that
\begin{equation}
F_{i,j}Q_{i+j}(\cB;\scrG(p))\, \big/\, F_{i-1,j+1}Q_{i+j}(\cB;\scrG(p)) \ \cong \  E_{i,j}^\infty \ .
\label{eq:extgroups}\end{equation}
If $\imath:\cB^i\hookrightarrow\cB$ denotes the inclusion of the $i$-skeleton in $\cB$, then the filtration groups
\begin{equation}
F_{i,j}Q_{i+j}(\cB;\scrG(p))={\rm im}\big(\imath_*:Q_{i+j}(\cB^i;\scrG(p))\to Q_{i+j}(\cB;\scrG(p)) \big)
\label{eq:filtgroups}\end{equation}
consist of branes in $\cB$ wrapping cycles supported in the $i$-skeleton, whereas the extension groups (\ref{eq:extgroups}) consist of branes in the $i$-skeleton which are not supported on the $i-1$-skeleton, i.e., $E_{i,j}^\infty$ consists of $i-1$-branes which carry no lower brane charges.

Let $\cM\subseteq \cB$ be a compact $i$-dimensional submanifold without boundary in $p:X\to\cB$ which defines a non-trivial cycle $[\cM]\in E_{i,j}^2$ in Eq.~(\ref{ss}). In the case that $[\cM]$ extends through the spectral sequence as a non-trivial element of $E_{i,j}^\infty$ and hence has a non-trivial lift to K-homology, there exists a stable D-brane $[\cW,F,\psi]$ wrapping $\cM$ on the $i$-skeleton of $\cB$ which carries no lower brane charges, i.e., $\psi(\cW)\subseteq\cB^i$. As the homotopy groups $\pi_j{\scrG}(p^{-1}(b))$ form stratified systems
of groups $\pi_j{\scrG}(p)$ over the background ${\cB}$, D-branes wrapping
spaces stratified fibered over hyperbolic orbifolds carry charges which induce new additive structures on
K-homology due to the stratification from $E_{i,j}^\infty$ in the
solution of the extension problem required to get the filtration
groups (\ref{eq:filtgroups}). Cycles with
$[\cM]\notin\ker(\dd^r_{i,j})$ correspond to anomolous
D-branes, while if $[\cM]\in{\rm im}(\dd^r_{i,j})$
for some $r$ then
the homology class $[\cM]$ can be lifted to K-homology but this lift is trivial as it vanishes in $E^\infty_{i,j}$: In this case there exists a D-brane wrapping $\cM$ in the $i$-th stratum $\cB_i$ with no lower brane charges, but this D-brane is unstable.

\bigskip

{\bf Examples. \ } 
{\bf (i)} \ Recalling that the differentials can be gleamed from singular homology (cf. Sect.~\ref{sec:homology}), it follows that $\dd^r_{i,j}=0$ for all $r$ even~\cite{Diaconescu} and the first non-trivial condition for a homology class $[\cM]\in E_{i,0}^2$ to survive to $E_{i,0}^\infty$ is given by
\begin{equation}
\dd^3_{i,0}[\cM] =0 \ .
\label{eq:d30}\end{equation}
This includes the Poincar\'e dual of the condition that the normal bundle $N\cM$ to $\cM$ in $\cB$ is a $Spin^\IC$ vector bundle~\cite{Diaconescu}, i.e., that $W_3(N\cM)=0$, where
$W_3(N\cM) \in H^3(\cM;\IZ)$ is the canonical integral lift of the third Stiefel-Whitney
class $w_3(N\cM) \in H^3(\cM;\IZ_2)$ of $N\cM$ which is the torsion class
defined as $W_3(N\cM)=\beta(w_2(N\cM))$, where $\beta:H^2(\cM;\IZ_2)\to
H^3(\cM;\IZ)$ is the Bockstein homomorphism. Applying the condition
(\ref{eq:d30}) to a D-brane $[\cW,F,\psi]$ wrapping $\cM$ requires the
normal bundle for the tangent bundle $T\cW$ with respect to the map $\psi:\cW\to\cB$ to be $Spin^\IC$; this is the real vector bundle $N\cW\to \cW$ such that $T\cW\oplus N\cW$ is isomorphic to the pullback by $\psi$ of a $Spin^\IC$ vector bundle over $\cB$ (recall that $\cW$ is already a $Spin^\IC$ manifold, i.e., a $Spin^\IC$ lift of $T\cW$ exists). Such a choice of K-orientation plays a crucial
role in determining the amount of supersymmetry which persists in the
hyperbolic orbifold backgrounds of Sect.~\ref{Hyper}.

{\bf (ii)} \ We can give a novel class of examples of unstable backgrounds which illustrates the destabilizing effects of the additional local fiber bundles over the strata of the hyperbolic orbifold $\cB$. Let $\scrG({\mathcal B}; p)$ be the homotopy cofiber of the map $ {\scrQ}({\mathcal B};
\scrG(p))\rightarrow \scrG(X)$ in the category of spectra,
which is also a loop spectrum. For each of the stratified systems of fiber bundles $p: X\rightarrow
{\mathcal B}$ there is a homotopy equivalence of loop spectra
${\scrG}(X) \cong {\scrQ}({\mathcal B}; \scrG(p))\times
\scrG({\mathcal B}; p), $ which is compatible with the
Atiyah-Hirzebruch type spectral sequence (\ref{ss}). Suppose that a stratified system of fiber bundles $p: X\rightarrow {\mathcal B}$ extends through the
spectral sequence as a non-trivial element of the homology groups
$E_{i,j}^2$. Then one can calculate the rational K-groups via~\cite[Thm. 2]{Farrell}, which asserts that if $\cX^\infty$ is aspherical with $Wh_i(\pi_1(\cX^\infty \times
S^1))\otimes \IQ = 0$ for all $i\in\IZ$, then $\scrG({\mathcal B};
p)\otimes \IQ = 0$. Hence the stratified system of fiber bundles
extends through the spectral sequence as a non-trivial
element of the homology groups, but if the homology is torsion-free it vanishes in $E_{i,j}^\infty$ and thus has a trivial lift to K-homology. In this case there exists a D-brane wrapping $\cB$, but this D-brane is unstable.

{\bf (iii)} \ For strongly virtually negatively curved groups $\Lambda$,
the groups $\Lambda_b = \pi_1(p^{-1}(b))$ for $b\in\cB$ are finite because they are
isomorphic to subgroups of $\tilde\Lambda$. As a consequence one can use
the wealth of results available for the algebraic K-theory of finite groups~\cite{Quillen,Borel} to formulate stability conditions for D-branes wrapping homology cycles. For example, if $\Lambda$ is strongly virtually negatively curved then
$K_n({\mathbb Z}{\Lambda}) = 0$ for all $n < - 1$, whereas 
$$ 
K_{-1}({\mathbb Z}{\Lambda})\cong
\lim_{{\stackrel{\scriptstyle \longrightarrow} {\scriptstyle L\in
\scrF({\Lambda})}}}\, K_{-1}({\mathbb Z}{L})
$$ 
is an abelian group finitely generated
by the images of $K_{-1}({\mathbb Z}{L})$, as $L$ varies over
the finite subgroups of $\Lambda$, under the map functorially induced
by the inclusion of $L$ into $\Lambda$.

{\bf (iv)} \ 
Let $\Lambda\subset G$ be a
discrete cocompact torsion-free subgroup as
above, and $\Gamma\subset G$ a maximal compact subgroup; let $X=G/\Gamma$. In this case we can use the orbifold stratified fiber bundle $p:X\to \cB$ to make the corresponding K-homology classes of D-branes on
$\cB=[\Lambda\setminus X]$ explicit. Given a
finite-dimensional unitary representation $\rho$ of the fundamental group $\Lambda$
there corresponds a locally homogeneous Chan-Paton vector bundle $\cE_\rho=\Lambda\setminus(X \times V_\rho)$ over $\cB$, where the
fiber $V_\rho$ of $\cE_\rho$ is the representation space of $\rho$ and
the $\Lambda$-action on $X\times V_\rho$ is defined by $\lambda\cdot(x,v):=
(\lambda\cdot x,\rho(\lambda)\, v)$ for $(\lambda,x,v)\in
\Lambda\times X\times V_\rho$. In the case that $X$ admits a
$G$-invariant $Spin^\IC$ structure, let $\mD_{\rho}$ be the
Dirac operator of $\cB$ acting on smooth sections of the homogeneous spinor bundle $\cS$ over
$\cB$ twisted by the vector bundle $\cE_\rho$; it is obtained by
projecting the Dirac operator $\mD$ of $X$ (which is $G$-invariant
and hence $\Lambda$-invariant) to $\cB$. As we discuss in
Sect.~\ref{KK-groups}, this defines an analytic K-homology class of $\cB$. The properties of this Dirac
operator can be studied by using the spherical harmonic analysis on
the rank one symmetric space $X$ developed by~\cite{Miatello1,Miatello2,BytsenkoWilliams} using Harish-Chandra's Plancherel density.

\section{Inclusion of $\boldsymbol H$-flux}
\label{KK-groups}

{\bf $\boldsymbol B$-fields and twisted K-theory. \ }
In the presence of a Neveu-Schwarz $B$-field whose curvature $H$ represents a non-trivial element
$[H]\in H^3(\cB; {\mathbb Z})$, the Chan-Paton vector bundles on D-branes should be replaced by suitable ``twisted" gauge bundles. Recall~\cite{Atiyah} that a twisted bundle $P_H\to \cB$ is given by a collection of locally defined bundles of Hilbert spaces $E_i\to U_i$ such that $P_H\big|_{U_i}\cong\IP(E_i)$ with respect to an open cover $\{U_i\}$ of the background $\cB$. The gluing functions between charts are realised by isomorphisms
$$
g_{ij}\,:\, E_i\big|_{U_i\cap U_j} \ \xrightarrow{ \ \sim \ } \ E_j\big|_{U_i\cap U_j} \ .
$$
On triple overlaps $U_i\cap U_j\cap U_k$ the composition $g_{ij}\circ
g_{jk}\circ g_{ki}$ is multiplication by a function $f_{ijk}:U_i\cap
U_j\cap U_k\to U(1)$; this data is equivalent to a lift
$g_{ij}:U_i\cap U_j\to U(\cH)$ of the $PU(\cH)$-valued one-cocycle $\tilde
g_{ij}:U_i\cap U_j\to PU(\cH)$ of the bundle $P_H\to \cB$, where $PU(\cH)=U(\cH)/U(1)$ is the group of projective unitary operators with
the topology induced by the norm topology on the unitary group
$U(\cH)$ of an infinite-dimensional separable Hilbert space $\cH$. The collection
$\{f_{ijk}\}$ satisfies the $U(1)$ cocycle condition $f_{ijk}\,
f_{jkl}^{-1}\, f_{kli}\, f_{lij}^{-1}=1$ on quadruple overlaps
$U_i\cap U_j\cap U_k\cap U_l$, and hence they define an integral \u{C}ech cocyle
$$
h_{ijkl}=\mbox{$\frac1{2\pi\ii}$} \, \big(\log f_{ijk}-\log f_{jkl}+\log f_{lij}\big)
$$
with \u{C}ech cohomology class $\big[\{h_{ijkl}\}\big]=[H]\in H^3(\cB; {\mathbb Z})$. In particular, on double overlaps the Hilbert bundles $E_i$ and $E_j$ differ by a line bundle $L_{ij}\to U_{i}\cap U_j$ with an isomorphism
$$
L_{ij}\otimes L_{jk}\cong L_{ik}
$$
on $U_i\cap U_j\cap U_k$ given by multiplication with $f_{ijk}$. The
set of line bundles $\{L_{ij}\}$ is called a bundle gerbe; the twisted
Chan-Paton bundles can then be regarded as bundle gerbe modules
associated to $\{L_{ij}\}$ and
the D-brane charges take values in the twisted K-theory defined as the
Grothendieck group of the additive category of lifting bundle gerbe modules~\cite{BCMMS}.

In this setting the $B$-field can be described as a connection on a $1$-gerbe whose Dixmier-Douady class is $[H]$, i.e., as a set of two-forms $B_i$ on $U_i$ satisfying the gluing conditions
$$
B_j-B_i=\mbox{$\frac1{2\pi\ii}$}\, \dd A_{ij}
$$
on $U_i\cap U_j$, where $A_{ij}$ is a connection on the line bundle $L_{ij}$ satisfying
$$
A_{ij}+A_{jk}+A_{ki}=\dd \log f_{ijk}
$$
on $U_i\cap U_j\cap U_k$; the collection of one-forms $\{A_{ij}\}$ specifies a set of connections $\nabla_i$ on the Hilbert bundles $E_i\to U_i$ with the gluing rules
$$
\nabla_i\big|_{U_i\cap U_j}=A_{ij}+\nabla_j\big|_{U_i\cap U_j} \ ,
$$
which defines a connection $\nabla$ on the twisted bundle $P_H \to\cB$ representing the $B$-field background $(\cB,B)$.
The $H$-flux is given locally by $H\big|_{U_i}=\dd B_i$ and thus has integral periods on~$\cB$.

These constructions all carry over to string orbifolds $\cB$ by
considering bundles on stacks, or equivalently on a presentation of
$\cB$ by a groupoid $\cG=(\cG_1\rightrightarrows\cG_0)$
(cf. Sect.~\ref{Hyper}). A $B$-field on a groupoid $\cG$ is a morphism
of groupoids $\tilde g:\cG\to PU(\cH)$ with respect to an open covering
$\{\cU_i\}$, where here the group $PU(\cH)$ is regarded as a groupoid $PU(\cH) \rightrightarrows\{1\}$. For local quotient groupoids, twisted K-theory is defined in~\cite{Freed} in terms
of bundles of Fredholm operators associated with such a twisted bundle of Hilbert spaces.

\bigskip

{\bf K-cycles and global anomaly cancellation. \ }
The corresponding twisted K-homology cycles $(\cW,F,\psi)$ encode the necessary data
that must be carried by any D-brane in an $H$-flux background in order
to fulfill the Freed-Witten anomaly cancellation
conditions~\cite{Szabo}, such as an $H$-twisted $Spin^{\mathbb
  C}$ structure and an ordinary complex vector bundle with $[F]\in
K^0(\cW)$~\cite{CareyWang}. Cancellation of global worldsheet
anomalies in the string theory sigma-model requires
\begin{equation}\label{eq:anomaly}
W_3(N\cW)+\psi^*[H]=0
\end{equation}
in $H^3(\cW;\IZ)$; the integral lift $W_3(N\cW)$ is the obstruction to
the existence of a $Spin^\IC$ structure on the stabilized normal bundle $N\cW$ over the worldvolume $\cW$. This
is a necessary condition for the homology class $\psi_*[\cW]$ to lift to
twisted K-homology; twisted $Spin^\IC$ structures (\ref{eq:anomaly})
are classified topologically by the integral cohomology $H^2(\cW;\IZ)$. To determine if a D-brane wrapping a
non-representable cycle $\cW$ carries K-theory charge, one must analyse the worldsheet open string theory, impose boundary conditions corresponding to a singular representative of the cycle, and then check for inconsistencies such as a failure of BRST invariance. Further instabilities can arise if there is a cycle $\psi':\cW'\to\cB$ such that $\cW$ is a codimension three submanifold of $\cW'$ satisfying the equation~\cite{Maldacena}
\begin{equation}
 W_3(N\cW'\,)+\psi'\,^* [H]={\rm Pd}_{\cW'}(\cW) \ ,
\label{eq:instanton}\end{equation}
where ${\rm Pd}_{\cW'}(\cW)$ is the Poincar\'e dual class of $[\cW]$ in $H^3(\cW';\IZ)$.

The original proposal of \cite{Witten,Kapustin,Bouwknegt} asserted that 
D-brane charges in Type~IIB string theory in a non-trivial $B$-field background are classified by the twisted K-theory of Rosenberg~\cite{Rosenberg}; in this framework the twisted bundles on the D-brane worldvolumes are bundles of Hilbert spaces associated with the infinite-dimensional locally trivial $C^*$-algebra bundle of compact operators $P_H(\cK):=P_H\times_{PU(\cH)}\cK$, where
$$
\cK=\lim_{\stackrel{\scriptstyle\longrightarrow}
{\scriptstyle n}}\, M_n(\IC)
$$
with the limit taken in the $C^*$-norm topology on the $n\times n$ matrix algebra $M_n(\IC)$. For our purposes we shall work mostly with the analytic formulation of K-homology as it is this version that is related to the definition of a twisted analogue of K-groups and the Kasparov map, which will allow us to use K-amenability results for Eilenberg-MacLane spaces.

\bigskip

{\bf Hyperbolic orbifolds. \ }
String backgrounds with hyperbolic orbifold factors become particularly important in the presence of non-vanishing $H$-flux. 
For example, solutions with brane charges can be found in a conformal field theory for the upper half three-space $H^3$ which is constructed as a WZW sigma-model based on the coset $SL(2,\IC)/SU(2)$. One can start with a brane solution with flux on $H^3$ which is a formal analog of the NS5-brane solution whereby the three-sphere is replaced by a hyperbolic space; its near-horizon geometry describes the background $AdS_3\times S^3\times H^3\times S^1$ with a linear dilaton in the time direction~\cite{Russo}. Such a construction leads to a $B$-field with imaginary components, but using S-duality it can be converted into a Ramond-Ramond two-form field with imaginary components which leads to a solution of Type~IIB$^*$ supergravity. As a result the conformal sigma-model is an exact solution of Type~II string theory to all orders in the $\alpha'$-expansion. Using U-duality one can also construct different D-brane solutions with time dependence; in order to accomodate a finite flux, the space $H^3$ should be replaced with a finite volume orbifold $H^3/\Lambda$.

In a manner akin to our ensuing analysis, a more algebraic perspective
on these solutions starts with the observation that the class of
Euclidean $AdS_3$ spaces we have considered here and in
Sect.~\ref{Hyper} are quotients of the real hyperbolic space $H^3$ by
a Schottky group, whose boundaries are compact oriented surfaces with
conformal structure. In a similar vein the boundary of the hyperbolic
plane $H^2$ at infinity is $\IR P^1$, and its global quotient by a
finite index subgroup $\Lambda$ of $G=PSL(2,\IZ)$ is a modular curve
$\cB=H^2/\Lambda$ which can be presented as the quotient
$\cB=(H^2\times G/\Lambda)/G$; the degree one homology classes of
$\cB$ can be regarded as classes in the cyclic cohomology of its
noncommutative boundary which is the crossed product $C^*$-algebra $C(\IR P^1\times
G/\Lambda)\rtimes G$ and is Morita equivalent to
$C(\IR P^1)\rtimes\Lambda$~\cite{Manin}. More generally, let $X$ be a
symmetric space of a real rank one semisimple Lie group, and let
$\Lambda\subset G$ be a discrete torsion-free subgroup; then the
geodesic boundary of $X$ has a $\Lambda$-equivariant decomposition $\partial X=\Omega_\Lambda\cup\Lambda_\infty$ where $\Lambda_\infty$ is the limit set of $\Lambda$, while the geometric boundary of the Poincar\'e half-space $H^n$ is $\IR^{n-1}_+$. If $\Lambda$ is convex cocompact (i.e., $X/\Lambda\cup\Omega_\Lambda$ is a compact manifold with boundary) then the orbit space $\cB=H^n/\Lambda$ can be regarded as the interior of a compact manifold with boundary, the Klein manifold $\cB\cup(\Omega_\Lambda/\Lambda)$, so that its geometric boundary at infinity is $\Omega_\Lambda/\Lambda$; see~\cite{Bytsenko1} for further details.

\bigskip

{\bf Kasparov theory. \ }
Recall~\cite{Wegge} that the reduced topological K-theory of the background $\cB$ can be defined as the K-theory $
\widetilde{K}^j(\cB)\cong {K}_{j}(C_0(\cB)),\,\, j = 0, 1
$ of the commutative $C^*$-algebra $C_0(\cB)$ of 
continuous complex-valued functions which vanish at infinity on
$\cB$. The definition of K-homology
involves classifying extensions of $C_0(\cB)$ by the algebra of compact
operators $\cK$ up to unitary equivalence \cite{Brown}. The set of
homotopy classes of operators defines the K-homology group
$K_0(\cB)$, and the duality with K-theory is provided by the natural
bilinear index pairing $ ([{\mathcal E}],[{\mathfrak D}])\mapsto {\rm Index}\,
{\mathfrak D}_{{\mathcal E}}\in {\mathbb Z}, $ where $[{\mathcal E}]\in K^0(\cB)$ and
${\mathfrak D}_{\mathcal E}$ denotes the action of the Fredholm operator
${\mathfrak D}$ on the Hilbert space ${\mathcal H}= L^2(\cB,{\mathcal E})$
of square-integrable sections of the vector bundle ${\mathcal E}\rightarrow
\cB$. If $\cB$ is a $Spin^\IC$ manifold then there is a Poincar\'e duality isomorphism between compactly supported K-theory and K-homology groups~\cite{Higson}.

In this context the KK-pairing appears to be the most natural framework. The group $K\!K_\sharp(A,B)$ is a bivariant version of
K-theory which depends on a pair of algebras $A$ and $B$; it defines a homotopy invariant bifunctor from the category of
separable $C^*$-algebras to the category of abelian groups which depends contravariantly on the algebra
$A$ and covariantly on the algebra $B$. It interpolates between K-theory and K-homology in the sense that 
$
K\!K^{\sharp}({A=\mathbb C}, B) = K_{\sharp}(B)
$ is the K-theory of $B$ while
$
K\!K^{\sharp}(A, B={\mathbb C}) = K^{\sharp}(A)
$ is the K-homology of $A$. Kasparov's analytic K-homology $K\!K^{\sharp}(C_0(\cB), \IC)$ is generated by unitary equivalence classes of (graded) Fredholm modules over $C_0(\cB)$ modulo an operator homotopy relation.
See \cite{Kasparov,Bytsenko} for a description of Kasparov's pairing
$
\otimes_D: K\!K(A,D)\times K\!K(D,B)\rightarrow K\!K(A,B)
$
and its properties. Applications of KK-theory to the classification of
D-branes can be found in e.g.~\cite{Periwal,Brodzki,Szabo}.

In the presence of a non-trivial $B$-field, the K-homology can be
defined analogously via the KK-groups of the noncommutative $C^*$-algebra
$C_0(\cB,P_H(\cK))$ of sections vanishing at infinity of the
associated bundle of compact operators $P_H(\cK)$. By identifying the
\u{C}ech homology of $\cB$ with its singular homology, this is the
unique (up to isomorphism) stable separable complex continuous trace
$C^*$-algebra with spectrum $\cB$ and Dixmier-Douady class $[H]\in
H^3(\cB;\IZ)$~\cite{Rosenberg,Atiyah}.

For a stack $\cB$ with groupoid presentation $\cG=(\cG_1\rightrightarrows\cG_0)$, it is convenient to work with a more algebraic definition~\cite{Tu}. For this, we pullback the central extension of groups
$$
1 \ \longrightarrow \ U(1) \ \longrightarrow \ U(\cH) \ \longrightarrow \
PU(\cH) \ \longrightarrow \ 1
$$
by the $B$-field $\tilde g:\cG\to PU(\cH)$ to get a central extension of the groupoid $\cG$ which is a principal $U(1)$-bundle $\tilde\cG_H \to\cG$ of groupoids with the same objects $\cG_0$; for a global quotient stack $[X/\Lambda]$ the gerbe is defined by a central extension $U(1)\to \tilde\Lambda_H\xrightarrow{\rm pr} \Lambda$ and the projection ${\rm pr}$ induces a map $[X/\tilde\Lambda_H]\to [X/\Lambda]$. Upon choosing a suitable Haar system on the groupoid, the twisted K-theory can be equivalently computed as the algebraic K-theory of the twisted groupoid $C^*$-algebra of $\cG$ defined as the subalgebra of $U(1)$-invariants of the convolution groupoid $C^*$-algebra $C^*(\tilde\cG_H)$~\cite{Tu}. Equivalently, as the gerbe associates a multiplier $\zeta\in H^2(\cG_1; U(1))$ on composable arrows, this algebra can be explicitly described as the twisted convolution $C^*$-algebra $C^*(\cG,\zeta)$ of the groupoid $\cG$; the two-cocycle $\zeta:\cG_2\to U(1)$ represents a class in $H^2(\cB;U(1))\cong H^3(\cB;\IZ)$ and we denote the corresponding $H$-flux by $[H_\zeta]$. A suitable variant of KK-theory in this setting is also defined by~\cite{Tu}.

\bigskip

{\bf K-amenable groups. \ }
Let $A,B$ be unital algebras for which there are elements
$\alpha\in K\!K(A\otimes B,{\mathbb C})$, $\beta\in K\!K({\mathbb
C},A\otimes B)$ with the property $\beta\otimes_{A}\alpha =
1_{B} \in K\!K(B,B)$,  $\alpha\otimes_{B}\beta = 1_{A} \in
K\!K(A,A).$ Then the algebras $A,B$ have canonically isomorphic K-theory and K-homology, and are said to be
KK-equivalent~\cite{Brodzki}. The notion of KK-equivalence can be used to introduce the concept of K-amenable
groups. As before, let $G$ be a connected Lie group and
${\Gamma}$ a maximal compact subgroup; we further assume that the symmetric space
$X= G/\Gamma$ has even dimension and admits a $G$-invariant ${Spin}^{\mathbb C}$ structure. The $G$-invariant
Dirac operator ${\mathfrak D }$ on $X$ is a first order self-adjoint elliptic differential operator acting on $L^2$-sections of the ${\mathbb Z}_2$-graded homogeneous bundle of spinors $\mathcal S$.
Consider the zeroth order pseudo-differential operator
${\mathcal D} = {\mathfrak D\, (1+{\mathfrak D}^2)^{-1/2}}$ acting on $\cH=L^2(X, {\mathcal S})$. The algebra $C_0(X)$ acts
on $\cH$ by multiplication operators. The group
$G$ acts on $C_0(X)$ and
$\cH$ by left translation, and ${\mathcal D}$ is $G$-invariant. Then the Fredholm module $({\mathcal D}, \cH)$ over $C_0(X)$ defines a canonical
{\it Dirac element} $\alpha_G \in K\!K_G(C_0(X), {\mathbb C})$, where $K\!K_G(A,B):= K\!K(A\rtimes G,B\rtimes G)$ for algebras $A$ and $B$ which admit an action of $G$ by automorphisms.
Kasparov shows~\cite{Kasparov} that there is a canonical {\it Mishchenko element}
$
\beta_G \in K\!K_G(\IC, C_0(X))
$
with the intersection products
$$
\alpha_G\otimes_{\mathbb C}\beta_G = 1_{C_0(X)} \ \in \  K\!K_G(C_0(X),C_0(X)) \ ,
$$
and
$$
\beta_G\otimes_{C_0(X)}\alpha_G=\gamma_G \ \in \ 
K\!K_G({\mathbb C}, {\mathbb C})
$$ 
where $\gamma_G$ is an idempotent element in
$K\!K_G({\mathbb C}, {\mathbb C})$.
For a semisimple Lie group $G$ or for $G={\mathbb R}^n$, a construction of the Mishchenko element $\beta_G$ can be found in \cite{Carey}. If the group $G$ is amenable, then $\gamma_G=1$. All solvable groups are amenable, while any non-compact semisimple Lie group is non-amenable. We therefore say that a Lie group $G$ is
K-amenable if $\gamma_G=1$.
\begin{proposition} \label{Proposition}
\begin{itemize}
\item[{\rm (a)}] Any amenable Lie group is K-amenable, and in particular any solvable Lie group is K-amenable.
\item[{\rm (b)}] The non-amenable groups ${{SO}}_0(n,1)$ and ${{SU}}(n,1)$ are K-amenable Lie groups {\rm \cite{Kasparov84,Julg}}.
\item[{\rm (c)}] The class of K-amenable groups is closed under the
operations of taking subgroups, and of direct and semidirect products
{\rm \cite{Cuntz}}.
\end{itemize}
\end{proposition}

\bigskip

{\bf Example: K-amenability in four dimensions. \ }
The problem of classifying geometries is one of the central problems in mathematics which also plays a fundamental role in constructing physical models.
Every one-dimensional manifold is either $S^1$ (closed) or ${\mathbb R}$ (open), with a unique topological piecewise linear smooth structure and orientation.
All complex curves of genus zero can be uniformized by rational functions, all those of genus one can be uniformized by elliptic functions,
and all those of genus larger than one can be uniformized by
meromorphic functions defined on proper open subsets of ${\mathbb C}$.
A complete solution to the uniformization problem in higher dimensions has not yet been achieved.
In three dimensions, the famous list of Thurston's eight
locally homogeneous spaces \cite{Thurston} can be organized into
compact stabilizer subgroups $\Gamma_b$ of $b\in \cB$ which are isomorphic to either
${SO}(3)$, ${SO}(2)$ or the trivial group; in the locally symmetric cases the isometry groups are all K-amenable~\cite{Carey}. The analogous list
of four-geometries (with connected isometry 
groups) can also be organized as in Tab.~1
\cite{Apanasov,Bytsenko}.
\begin{table}\label{Table2}
\begin{center}
\begin{tabular}
{l l}
Table~1. Four-geometries
\\
\\
\hline
\\
Stabilizer subgroup $\Gamma$ & Background $\cB=\Lambda\setminus G/\Gamma$ \\
\\
\hline
\\
${SO}(4)$ & ${S}^4,\,\,{\mathbb R}^4,\,\, {H}^4$ \\
${U}(2)$ & ${\mathbb C} P^2,\,{\mathbb C}{H}^2$ \\
${SO}(2)\times {SO}(2)$ & ${S}^2\times {\mathbb R}^2,\,\,
{S}^2\times
{S}^2,\,\,{S}^2\times {H}^2,\,\, {H}^2\times
{\mathbb R}^2,\,\,{H}^2\times {H}^2$ \\
${SO}(3)$ & ${S}^3\times {\mathbb R},\,\, {H}^3\times
{\mathbb R}$ \\
${SO}(2)$ & ${N}il^3\times
{\mathbb R},\,\,{\widetilde{{PSL}}}(2,{\mathbb R})\times {\mathbb R},\,\,{S}ol^4$ \\
$U(1)$ & $\mathbb{F}^4$ \\
{\rm trivial} & ${N}il^4,\,\, {S}ol^4_{m,n} , \ {S}ol^4_1$ \\
\\
\hline
\end{tabular}
\end{center}
\end{table}
Here we have the four irreducible four-dimensional Riemannian symmetric spaces: sphere ${ S}^4$,
hyperbolic space ${H}^4$, complex projective space ${\mathbb C}
P^2$ and complex hyperbolic space ${\mathbb C} {H}^2$, the nilpotent and solvable Lie groups (including
$Sol_{m,m}^4={S}ol^3\times {\mathbb R}$), and the space $\mathbb{F}^4$ with isometry group $\IR^2\rtimes PSL(2,\IR)$.
\begin{corollary} The isometry groups $G$ of the locally symmetric backgrounds $\cB=\Lambda\backslash X$ listed in Tab. 1 are all K-amenable.
\label{Corollary}\end{corollary}
To establish this result we need to prove that $\gamma_G=1$ for each of the four-geometries occuring in Tab. 1. Let us consider just a few representative examples, see \cite{Bytsenko} for further details:
\begin{itemize}
\item[---] $G = {\mathbb R}^4\rtimes SO(4), \, X = {\mathbb R}^4$:
By Prop. \ref{Proposition}, $\gamma_G = 1$ since ${\mathbb R}^4$ and $SO(4)$ are amenable, and 
so is their semidirect product.
\item[---] $G = SO_0(4,1) , \,X = H^4$: $\gamma_G = 1$ by Prop. \ref{Proposition}. 
\item[---] $G = SU(3), \, X =
{\mathbb C}P^2\cong U(3)/(U(1)\times U(2))\cong
SU(3)/S(U(1)\times U(2))$: $\gamma_G = 1$ by Prop.
\ref{Proposition}. 
\item[---] $X = {\mathbb C}H^2$: The
four-geometry ${\mathbb C}H^2$ is a K\"ahlerian symmetric space and carries a complex structure, hence $\gamma_G = 1$. 
\item[---] \ $X = H^2\times {\mathbb R}^2, H^2\times H^2, H^3\times {\mathbb
R}$: By Prop. \ref{Proposition} one has $\gamma_G = 1$ since these isometry groups are direct products of
K-amenable groups.
\end{itemize}

\bigskip

{\bf Twisted K-groups. \ }
Let $A$ be an algebra
admitting an action of a lattice $\Lambda\subset G$ by automorphisms.
The crossed product algebra $\big(A\otimes C_0(X) \big) \rtimes \Lambda$
is Morita equivalent to the algebra $C_0(\cB, {\mathcal E})$ of continuous sections
vanishing at infinity of the flat $A$-bundle defined by the quotient
${\mathcal E}:= \Lambda\setminus (X\times A) \rightarrow \cB$ with the diagonal action of $\Lambda$ on $X\times A$~\cite{Carey}. One has the following version of the Thom isomorphism theorem for the K-theory of $C^*$-algebras.
\begin{theorem}
{\rm \cite{Kasparov}}\label{Th1}
If $G$ is K-amenable, then
$(A\rtimes\Lambda)\otimes C_0(X)$ and $\big( A\otimes C_0(X) \big) \rtimes
\Lambda$ have the same K-theory.
\end{theorem}
It follows that when $G$ is K-amenable, the algebras $(A\rtimes\Lambda)\otimes C_0(X)$ and $C_0(\cB, {\mathcal E})$
have the same K-theory so that 
\begin{equation*}
K_\sharp(C_0(\cB; {\mathcal E})) \cong K_{\sharp + \dim(X)}
(A\rtimes\Lambda) \ .
\end{equation*}
Let $\zeta \in H^2(\Lambda; U(1))$ be a multiplier on $\Lambda$, i.e.,
a normalized $U(1)$-valued group two-cocycle on $\Lambda$.
\begin{theorem} 
{\rm \cite{Carey}}\label{main}
If $G$ is K-amenable, then
\begin{equation}
K_\sharp(C^*(\Lambda,\zeta)) \cong K^{\sharp + \dim(X)}(\cB,
[H_\zeta])\ ,
\label{K-am}
\end{equation}
where $K^{\sharp}(\cB,
[H_\zeta])$ is the twisted K-theory of the continuous trace $C^*$-algebra
$C_0(\cB,P_{H_\zeta}(\cK))$ with Dixmier-Douady invariant $[H_\zeta]\in H^3(\cB;\IZ)$.
\end{theorem}
To better understand this result, first suppose that $A={\mathbb C}$ with $\Lambda$ acting trivially.
When $\gamma_G=1$, Thm.~\ref{Th1} implies that the algebras $({\mathbb C}\rtimes \Lambda)\otimes C_0(X)$ and $C_0(\cB, {\mathcal E})$ have the same K-theory, where $\cE$ is the trivial complex line bundle over $\cB$; it follows that $K_\sharp(C^*(\Lambda)) \cong K^{\sharp + \dim(X)}(\cB)$. Now suppose that $\zeta\in H^2(\Lambda; {U}(1))$. If $G$ is K-amenable, then Thm.~\ref{Th1} and the Packer-Raeburn stabilization trick~\cite{Packer} imply that 
$
(A\rtimes_\zeta \Lambda)\otimes C_0(X)
$ and 
$C_0(\cB, {\mathcal E}_\zeta)
$
have the same K-theory, where ${\mathcal E}_\zeta:= \Lambda\setminus \big( X\times (A\otimes{\cK})\big) /\Lambda\rightarrow \cB$ with $\cK$ the algebra of compact operators.
Since by definition the twisted K-theory $K^\sharp(\cB, [H_\zeta])$ is the K-theory of the continuous trace $C^*$-algebra
$C_0(\cB, {\mathcal E}_\zeta)$, Eq. (\ref{K-am}) thus follows. We may
regard this result as saying that twisted D-brane charges in the
$H$-flux background are the same as those of D-branes in a
noncommutative deformation of the locally symmetric space $\cB$
induced by a generating $B$-field. 

\bigskip

{\bf Examples. \ } {\bf (i)} \ General theorems regarding the K-groups of $C^*$-algebras are obtained in~\cite{Elliott,Bellissard,Connes},
while the analysis of Baum-Connes type conjectures concerning the K-theories of twisted group $C^*$-algebras is carried out in \cite{Marcolli,Mathai}. Collecting these results in the generic case one finds
$
K_\sharp(C^*(\mathbb Z^n,\zeta)) \cong K_\sharp(C^*(\mathbb Z^n)) \cong K^\sharp(T^n)
$
for any multiplier $\zeta$ on $\mathbb Z^n$. The twisted group
$C^*$-algebras $C^*(\mathbb Z^n,\zeta)$ are the well-known noncommutative $n$-tori.

{\bf (ii)} \ When $\Lambda = \Lambda_g$ is the fundamental group of a
Riemann surface $\Sigma_g$ of genus $g>0$, the
Dixmier-Douady class $[H_{\zeta}]$ is trivial and we get \cite{Carey}
$
K_0(C^*({\Lambda}_g,\zeta)) \cong
K^{0}(\Sigma_g) \cong {\mathbb Z}^2,\,
K_1(C^*({\Lambda}_g,\zeta)) \cong
K^{1}(\Sigma_g) \cong {\mathbb Z}^{2g}
$
for any multiplier $\zeta$ on $\Lambda_{g}$. The twisted group
$C^*$-algebra $C^*(\Lambda_g,\zeta)$ is called a {noncommutative Riemann surface}.

{\bf (iii)} \ Since the symmetric space $X=G/\Gamma$ is contractible and the group $\Lambda$ acts freely on $X$, one has $H_i(\Lambda; A)= H_i(\cB; A)$ and
$H^i(\Lambda; A)= H^i(\cB; A)$ for any trivial $\Lambda$-module $A$ where the groups on the left-hand sides are Eilenberg-MacLane homology and cohomology groups. As an explicit example, let $G= {\mathbb R}^n$, $\Lambda = {\mathbb Z}^n$, and
$\Gamma=\{0\}$. In this case $\cB= \Lambda\backslash G/\Gamma= {T}^n$ is an $n$-torus with
\begin{equation*}
H_i({T}^n; {\mathbb Z})=H^i({T}^n; {\mathbb Z}) =
H^i({\IZ}^n; {\mathbb Z})= {\mathbb Z}^{{{n\choose i}}} \ .
\end{equation*}
Then for $j=0,1$ one has
\begin{equation*}
K_j(T^n)=K^j(T^n)= \bigoplus_{i=0}^{\lfloor\frac{n}{2}\rfloor}\, H^{2i+j}(T^n;\IZ) = \prod_{i=0}^{\lfloor\frac{n}{2}\rfloor}\,{\mathbb Z}^{{n\choose 2i+j}} = {\mathbb Z}^{2^{n-1}} \ .
\end{equation*}
The power
$2^{n-1}$ gives the expected multiplicity of brane charges arising from wrapping all higher stable D-branes on various cycles of the torus ${T}^n$~\cite{Olsen}.

\bigskip

{\bf Eilenberg-MacLane spectrum functor. \ }
One of the main results of \cite{Carey} says that for lattices in K-amenable Lie groups the reduced and unreduced twisted group
$C^*$-algebras have canonically isomorphic K-theory.
If $\zeta\in H^2(\Lambda, {U}(1))$ is a multiplier on a lattice $\Lambda$ in a K-amenable Lie group $G$, then the canonical morphism $C^*(\Lambda, \zeta)\rightarrow
C_r^*(\Lambda, \zeta)$ induces an isomorphism
\begin{equation*}
K_\sharp(C^*(\Lambda, \zeta))\cong K_\sharp(C_r^*(\Lambda, \zeta))\,.
\end{equation*}
If moreover $\dim(X) = 3$ and $\Lambda$ is a uniform
lattice in $G$, then for $j =0,1$ one has
\begin{equation}
K_j(C^*_r(\Lambda, \zeta)) \cong K_j(C^*_r(\Lambda)) \cong
K^{j+1}(\cB)\ .
\label{K11}
\end{equation}
To show this, we use Thm. \ref{main} to get
$
K_j(C^*_r(\Lambda)) \cong K^{j+\dim(X)}(\cB)
$
for $j=0,1$.
By the Packer-Raeburn stabilization trick \cite{Packer}, $C_r^*(\Lambda,\zeta)$ is Morita
equivalent to $\cK\rtimes\Lambda$ and since $G$ is K-amenable
$(\cK\rtimes\Lambda) \otimes C_0(X)$ is Morita equivalent to
$C_0(\cB, {\mathcal E}_\zeta)$, where as before ${\mathcal E}_\zeta$ is a locally trivial bundle of $C^*$-algebras
over $\cB$ with fiber $\cK$. The Dixmier-Douady invariant is an element
$
[H_\zeta] \in
H^3(\cB; \mathbb{Z})
\cong H^3(\Lambda;\mathbb{Z})\,.
$
If $\cB$ is not orientable, then $H^3(\cB; {\mathbb Z}) = \{0\}$; therefore
$[H_\zeta] =0$ and Eq. (\ref{K11}) holds.
On the other hand, when $\cB$ is orientable one has $H^3(\cB; {\mathbb Z}) \cong\IZ$ but
$[H_\zeta]=0$ for all $\zeta\in H^2(\Lambda, U(1))$ \cite{Carey}, and so
$C_0(\cB, {\mathcal E}_\zeta)$ is Morita equivalent to $C_0(\cB)$; in this case we have again Eq. (\ref{K11}).

Let $\underline{K}(\Lambda, n)$ be an
Eilenberg-MacLane space, i.e., the unique space (up to homotopy) whose
homotopy groups are $\pi_m(\,\underline{K}(\Lambda,n))=0$ for all
$m\neq n$ and $\pi_n(\,\underline{K}(\Lambda,n))=\Lambda$; for
example one has
$\underline{K}(\IZ^n,1)=T^n$ and $\underline{K}(\IZ,2)=\IC P^\infty\cong \underline{B}U(1)$
is the classifying space of $U(1)$-bundles, while $\underline{K}(\IZ,2)=PU(\cH)$ and
$\underline{K}(\IZ,3)= \underline{B}PU(\cH)$ is the classifying space of twisted bundles. The spaces $ \{\,\underline{K}(\Lambda, n)\}_{n\geq0}$ fit together via the loop space functor to form the Eilenberg-MacLane spectrum
$$
\cdots \ \xleftarrow{ \ \Omega \ } \ \underline{K}(\Lambda,n) \ \xleftarrow{ \ \Omega \ } \ \underline{K}(\Lambda,n+1) \ \xleftarrow { \ \Omega \ } \ \cdots
$$
which we can regard as a functor from the category of abelian groups to the category of loop spectra.

Any compact connected locally symmetric background $\cB$ is an Eilenberg-MacLane space $\underline{K}(\Lambda,1)$. Conversely, if $\underline{K}(\Lambda,1)$ is a locally
symmetric manifold then it is of the form $\Lambda\backslash G/\Gamma$ where $G$
is a connected Lie group, $\Gamma$ is a maximal compact subgroup of
$G$ and $\Lambda$ is a uniform lattice in $G$. In~\cite{Carey} it is conjectured that Eq. (\ref{K11}) is still valid without the assumption of local symmetry:
If $\cB$ is a connected compact three-manifold which is an Eilenberg-MacLane space with fundamental group $\Lambda$, then for any multiplier
$\zeta\in H^2(\Lambda, U(1))$ on $\Lambda$ one has isomorphisms
$
K_j(C_r^*(\Lambda,\zeta)) \cong K_j(C_r^*(\Lambda)) \cong K^{j+1}(\cB)$ for $j=0,1.
$
This statement is extended to the four-manifolds of Tab.~1 in \cite{Bytsenko}.

\bigskip

{\bf Brane stability conditions. \ }
The problem of stability of the supergravity solutions discussed above can be embedded into the framework of Sect.~\ref{stability}. Given a stratified system of fiber bundles $p:X\to \cB$ over a hyperbolic
orbifold which supports a non-trivial $B$-field $P_H\to \cB$, one can repeat
the topological classification of stable branes in terms of Quinn homology groups
$Q_i(\cB;\scrG(p))$ by using spectra suitable to the appropriate
twisted homology theory; in this case we assume that $\cB$ can be
covered by suitable neighborhoods $U_i$ of the strata $\cB_i$. For
D-branes, we identify the Bott spectrum $\scrK$ with the iterated loop
spectrum of the space of Fredholm operators and choose a spectrum-valued functor of spaces $\scrG$ such that $\scrG(X)$ is homotopy equivalent to the loop spectrum
which is the pullback by $p$ of the associated bundle of
K-homology spectra $P_H(\scrK):= P_H\times_{PU(\cH)}\scrK$. In this case
anomaly cancellation (\ref{eq:anomaly}) and instability
(\ref{eq:instanton}) are consistent as before with the homology of the
differential $\dd^3_{i,0}$ in the spectral sequence extension
problem~\cite{Maldacena}, but unfortunately not very much is
known about the higher differentials in general.
 Other branes can be obtained by duality.

A somewhat simpler analysis can be carried out using Eilenberg-MacLane spaces.
If $A$ is a stratified system of abelian groups over $\cB\supset
\cB^i\supset \cB^{i-1}\supset\cdots$ as in
Sect.~\ref{stability},
we can apply the Eilenberg-MacLane spectrum functor to its fibers.
For each $b\in\cB$ with $b\in \cB^i$ this gives fiber bundles
\begin{equation*}
\underline{K}(A(b),n) \ \longrightarrow \ \underline{K}(A_i,n) \ 
\longrightarrow \ U_i \ .
\end{equation*}
The homomorphisms $\theta_{ij}:A_i\to A_j$ define
fiber maps over $U_i\cap U_j$ which can be used to define a topology
on the disjoint union
$\underline{K}(A,n) := \coprod_{i\geq1}\, \underline{K}(A_i,n)$. Applying the Eilenberg-Steenrod additivity
axiom to any sequence of homology functors $H_n$ gives
\begin{equation*}
H_\sharp(\,\underline{K}(A,n)) \cong \bigoplus_{i\geq1}\,
H_\sharp(\,\underline{K}(A_i,n)) \ .
\end{equation*}
If each $\cB^i$ is an absolute neighborhood retract (i.e., $\cB^i$ has a neighborhood which retracts it when embedded
as a closed subspace of a normal space), 
then $\underline{K}(A,n)\rightarrow \cB$ is a stratified system of
fiber bundles which fit together to give a spectral sheaf $\scrG_A$ over $\cB$. The corresponding homology
spectrum is denoted 
$
{\scrQ}(\cB; A) :=
{\scrQ}(\cB; \scrG_A),
$
and as before the Quinn homology groups are computed from the homotopy of
this spectrum as $Q_j(\cB; A) = \pi_j
{\scrQ}(\cB; A)$. 

By Sect.~\ref{sec:homology} these homology groups may be computed from an Atiyah-Hirzebruch type spectral
sequence with $E^2$ term
$$
E_{i,j}^2\cong\bigoplus_{k\geq1}\, H_i(\cB;\pi_j(\scrG_{A_k})) \ .
$$
As in Sect.~\ref{stability}, if $\underline{K}(A,n)$
extends through this spectral
sequence as a non-trivial element of all homology groups, then it can
pass to a non-trivial element of $E^\infty_{i,j}$ and hence have a
non-trivial lift to K-homology. Then the twisted additive structure of
the K-homology cycles will encode the lower brane charges that must be
carried by any D-brane which wraps the Eilenberg-MacLane space
$\underline{K}(A,n)$.

This construction provides a novel mechanism for stability of the hyperbolic orbifold background $\cB$, wherein it can be extended via a stratified system of groups to a non-trivial element of K-homology. As the homotopy groups of $\underline{K}(A,n)$ are mostly trivial, this imposes much more stringent constraints on the allowed stable branes in such $B$-field backgrounds. A model for each space $\underline{K}(A_i,n)$ can be obtained by taking a presentation of $A_i$ and the smash product $S^n\wedge\cdots\wedge S^n$, with one $n$-sphere for each generator of $A_i$, and then attaching $n+1$-cells for each relation. If $n=1$ and $A_i$ has torsion, then no finite-dimensional CW-complex can model $\underline{K}(A_i,1)$ as then its homology is torsion in each odd degree by the Dold-Thom theorem; for example $\underline{K}(\IZ_m,1)$ can be modelled by the infinite Lens space $S^\infty/\IZ_m$.

\subsection*{Acknowledgments}

We are grateful to Alan Carey and Masud Chaichian for helpful discussions and suggestions. AAB would like to acknowledge the Conselho Nacional
de Desenvolvimento Cient\'ifico e Tecnol\'ogico (CNPq, Brazil) and Funda\c cao Araucaria
(Parana, Brazil) for financial support. The support of the Academy of Finland under the
Projects No. 136539 and 272919 is gratefully acknowledged.
The work of RJS was partially supported by the Consolidated Grant
ST/L000334/1 from the
UK Science and Technology Facilities Council. Part of this work was
carried out
while RJS was visiting the Hausdorff Research Institute for
Mathematics in Bonn during the 2014 Trimester Program ``Noncommutative
Geometry and its Applications''; he would like to thank Alan Carey,
Victor Gayral, Matthias Lesch, Walter van Suijlekom and Raimar
Wulkenhaar for the invitation, and all the staff at HIM for the warm
hospitality.


\begin{thebibliography}{99}

\bibitem{Dixon}
  L.~J.~Dixon, J.~A.~Harvey, C.~Vafa and E.~Witten,
  {\it Strings on Orbifolds},
  Nucl.\ Phys.\ B {\bf 261} (1985) 678--686.
  
\bibitem{Sharpe}
E.~R.~Sharpe,
  {\it String Orbifolds and Quotient Stacks},
  Nucl.\ Phys.\ B {\bf 627} (2002) 445--505;
  [arXiv:hep-th/0102211].
  
\bibitem{Pantev}
  T.~Pantev and E.~R.~Sharpe,
  {\it String Compactifications on Calabi-Yau Stacks},
  Nucl.\ Phys.\ B {\bf 733} (2006) 233--296;
  [arXiv:hep-th/0502044].

\bibitem{FigueroaO'Farrill}
  J.~Figueroa-O'Farrill and N.~Hustler,
  {\it The Homogeneity Theorem for Supergravity Backgrounds},
  JHEP {\bf 1210} (2012) 014;
  [arXiv:hep-th/1208.0553].

\bibitem{Quinn}
F. Quinn, {\it Ends of Maps II}, Invent. Math. {\bf 68} (1982)
353--424.

\bibitem{Adem}
A.~Adem, J.~Leida and Y.~Ruan,
{\sl Orbifolds and Stringy Topology} (Cambridge University Press,
Cambridge, 2007).

\bibitem{Heinloth}
J.~Heinloth, 
{\it Notes on Differentiable Stacks},  in: {\sl Mathematisches
  Institut Georg-August-Universit\"at G\"ottingen Seminars Winter Term
  2004/2005}, ed. Yu.~Tschinkel (Universit\"atsdrucke G\"ottingen,
2005) 1--32.

\bibitem{Hughes}
B.~Hughes,
{\it Stratified Path Spaces and Fibrations},
Proc. Roy. Soc. Edinburgh A {\bf 129} (1999) 351--384.

\bibitem{Freed}
D.~S.~Freed, M.~J.~Hopkins and C.~Teleman,
{\it Loop Groups and Twisted K-Theory I},
J. Topol. {\bf 4} (2011) 737--798;
[arXiv:math.AT/0711.1906].

\bibitem{BF}
P.~Breitenlohner and D.~Z.~Freedman, {\it Stability in Gauged
Extended Supergravity}, Ann. Phys. {\bf 144} (1982) 249--281.

\bibitem{FY}
Y. Fujii and K. Yamagishi, {\it Killing Spinors on Spheres and
Hyperbolic Manifolds}, J. Math. Phys. {\bf 27} (1986) 979--981.

\bibitem{LP}
H.~Lu, C.~N.~Pope and P.~K.~Townsend, {\it Domain Walls from Anti-de
Sitter Spacetime}, Phys. Lett. B {\bf 391} (1997) 39--46;
[arXiv:hep-th/9607164].

\bibitem{Lu}
H.~Lu, C.~N.~Pope and J.~Rahmfeld, {\it A Construction of Killing
Spinors on $S^n$}, J. Math. Phys. {\bf 40} (1999) 4518--4526;
[arXiv:hep-th/9805151].

\bibitem{Russo}
A. Kehagias and J. G. Russo, {\it Hyperbolic Spaces in String and
M-Theory}, JHEP {\bf 0007} (2000) 027; [arXiv:hep-th/0003281].

\bibitem{kach}
S.~Kachru and E.~Silverstein, {\it $4d$ Conformal Field Theories and
Strings on Orbifolds}, Phys. Rev. Lett. {\bf 80} (1998) 4855--4858;
[arXiv:hep-th/9802183].

\bibitem{NV}
A.~Lawrence, N.~A.~Nekrasov and C.~Vafa, {\it On Conformal Theories
in Four Dimensions}, Nucl. Phys. B {\bf 533} (1998) 199--209;
[arXiv:hep-th/9803015].

\bibitem{Friedrich}
T. Friedrich, {\it Dirac Operators in Riemannian Geometry},
Grad. Stud. Math. {\bf 25} (1997) 1--193.

\bibitem{Talbert}
R.~N.~Talbert, {\it An Isomorphism between Bredon and Quinn Homology}, Forum Math.
{\bf 11} (1999) 591--616.

\bibitem{Bousfield}
A. K. Bousfield and D. M. Kan, {\it Homotopy Limits, Completions
and Localizations}, Lect. Notes Math. {\bf 304} (1972) 1--348.

\bibitem{Whitehead}
G. W. Whitehead, {\it Generalized Homology Theories}, Trans.
Amer. Math. Soc. {\bf 102} (1962) 227--283.

\bibitem{James}
I.~M.~James,
{\it Ex-Homotopy Theory},
Illinois J. Math. {\bf 15} (1971) 324--337.

\bibitem{Whitehead2}
G. W. Whitehead, {\it Elements of Homotopy Theory},
Grad. Texts Math. {\bf 61} (1988) 1--736.

\bibitem{Farrell}
F. T. Farrell and L. E. Jones, {\it Algebraic K-Theory of Spaces
Stratified Fibered over Hyperbolic Orbifolds}, Proc. Natl. Acad.
Sci. USA {\bf 83} (1986) 5364--5366.

\bibitem{Davis}
J.~F.~Davis and W.~L\"uck,
{\it Spaces over a Category and Assembly Maps in Isomorphism Conjectures in K- and L-Theory}, K-Theory {\bf 15} (1998) 201--252.

\bibitem{SzaboValentino}
  R.~J.~Szabo and A.~Valentino,
 {\it Ramond-Ramond Fields, Fractional Branes and Orbifold Differential K-Theory},
  Commun.\ Math.\ Phys.\ {\bf 294} (2010) 647--702;
  [arXiv:hep-th/0710.2773].
  
\bibitem{Short}
A. A. Bytsenko, M. Chaichian and M. E. X. Guimar\~aes, {\it D-Branes on Spaces Stratified 
Fibered Over Hyperbolic Orbifolds}, Int. J. Mod. Phys. A {\bf 29} (2014) 1450137; [arXiv:hep-th/1408.1071].

\bibitem{Reis}
R. M. G. Reis and R. J. Szabo, {\it Geometric K-Homology of Flat
D-Branes}, Commun. Math. Phys. {\bf 266} (2006) 71--122; [arXiv:hep-th/0507043].

\bibitem{Diaconescu}
  D.-E.~Diaconescu, G.~W.~Moore and E.~Witten,
  {\it $E_8$ Gauge Theory and a Derivation of K-Theory from M-Theory},
  Adv.\ Theor.\ Math.\ Phys.\ {\bf 6} (2003) 1031--1134;
  [arXiv:hep-th/0005090].

\bibitem{Quillen}
D. Quillen, {\it Finite Generation of the Groups $K_i$ of Rings of
Algebraic Integers},
Lect. Notes Math. {\bf 341} (1973) 179--198.

\bibitem{Borel}
A. Borel, {\it Stable Real Cohomology of Arithmetic Groups}, Ann.
Sci. \'Ecole Norm. Sup. {\bf 7} (1974) 235--272.

\bibitem{Miatello1}
R.~Miatello,
{\it On the Plancherel Measure for Linear Lie Groups of Rank One},
Manuscr. Math. {\bf 29} (1979) 249--276.

\bibitem{Miatello2}
R.~Miatello,
{\it The Minakshisundaram-Pleijel Coefficients for the Vector-Valued
  Heat Kernel on Compact Locally Symmetric Spaces of Negative Curvature},
Trans. Amer. Math. Soc. {\bf 260} (1980) 1--33.

\bibitem{BytsenkoWilliams}
A.~A.~Bytsenko and F.~L.~Williams,
{\it Asymptotics of the Heat Kernel on Rank One Locally Symmetric
  Spaces},
J. Phys. A {\bf 32} (1999) 5773--5780;
[arXiv:math.SP/9804115].

\bibitem{Atiyah}
M.~F. Atiyah and G.~B. Segal, {\it Twisted K-Theory}, Ukr. Math. Bull. {\bf 1} (2004) 291--334; [arXiv:math.KT/0510674].

\bibitem{BCMMS}
  P.~Bouwknegt, A.~L.~Carey, V.~Mathai, M.~K.~Murray and D.~Stevenson,
  {\it Twisted K-Theory and K-Theory of Bundle Gerbes},
  Commun.\ Math.\ Phys.\  {\bf 228} (2002) 17--49;
  [arXiv:hep-th/0106194].

\bibitem{Szabo}
R.~J.~Szabo,
{\it D-Branes and Bivariant K-Theory},
Keio COE Lect. Ser. Math.
Sci. {\bf 1} (2013) 131--175;
[arXiv:hep-th/0809.3029].

\bibitem{CareyWang}
  A.~L.~Carey and B.-L.~Wang,
  {\it Riemann-Roch and Index Formulae in Twisted K-Theory},
  Proc. Symp. Pure Math. {\bf 81} (2010) 95--131;
  [arXiv:math.KT/0909.4848].
  
\bibitem{Maldacena}
  J.~M.~Maldacena, G.~W.~Moore and N.~Seiberg,
  {\it D-Brane Instantons and K-Theory Charges},
  JHEP {\bf 0111} (2001) 062;
  [arXiv:hep-th/0108100].

\bibitem{Witten}
E. Witten, {\it D-Branes and K-Theory}, JHEP {\bf 9812} (1998)
019; [arXiv:hep-th/9810188].

\bibitem{Kapustin}
A.~Kapustin, {\it D-Branes in a Topologically Nontrivial $B$-field},
Adv. Theor. Math. Phys. {\bf 4} (2000) 127--154; [arXiv:hep-th/9909089].

\bibitem{Bouwknegt}
P. Bouwknegt and V. Mathai, {\it D-Branes, $B$-Fields and Twisted
K-Theory}, JHEP {\bf 0003} (2000) 007; [arXiv:hep-th/0002023].

\bibitem{Rosenberg}
J. Rosenberg, {\it Continuous Trace Algebras from the Bundle
Theoretic Point of View}, J. Austral. Math. Soc. {\bf 47} (1989)
368--381.

\bibitem{Manin}
Yu. I. Manin and M. Marcolli, {\it Holography Principle and
Arithmetic of Algebraic Curves}, Adv. Theor. Math. Phys. {\bf 5}
(2002) 617--650; [arXiv:hep-th/0201036].

\bibitem{Bytsenko1}
A. A. Bytsenko, M. E. X. Guimar\~aes and J. A. Helayel-Neto, {\it Hyperbolic Space Forms and Orbifold 
Compactification in M-Theory}, PoS WC {\bf 2004} (2004) 017; [arXiv:hep-th/0502031].

\bibitem{Wegge}
N. E. Wegge-Olsen, {\sl K-Theory and $C^*$-Algebras} (Oxford University Press, Oxford, 2000).

\bibitem{Brown}
L. Brown, R.~G. Douglas and P. Fillmore, {\it Extensions of
$C^*$-Algebras and K-Homology}, Ann. Math. {\bf 105} (1977)
265--324.

\bibitem{Higson}
N. Higson and J. Roe, {\sl Analytic K-Homology}
(Oxford University Press, Oxford, 2000).

\bibitem{Kasparov}
G. G. Kasparov, {\it K-Theory, Group $C^*$-Algebras and Higher
Signatures}, London Math. Soc. Lect.
Notes {\bf 226} (1995) 101--146.

\bibitem{Bytsenko}
A. A. Bytsenko, {\it Homology and K-Theory Methods for Classes of
Branes Wrapping Nontrivial Cycles}, J. Phys. A
{\bf 41} (2008) 045402; [arXiv:hep-th/0710.0305].

\bibitem{Periwal}
V. Periwal, {\it D-Brane Charges and K-Homology}, JHEP {\bf 0007}
(2000) 041; [arXiv:hep-th/0006223].

\bibitem{Brodzki}
J. Brodzki, V. Mathai, J. Rosenberg and R. J. Szabo, 
{\it D-Branes, RR-Fields and Duality on Noncommutative Manifolds},
Commun. Math. Phys. {\bf 277} (2008) 643--706; [arXiv:hep-th/0607020].

\bibitem{Tu}
  J.~L.~Tu, P.~Xu and C.~Laurent,
  {\it Twisted K-Theory of Differentiable Stacks},
  Ann. Sci. \'Ecole Norm. Sup. {\bf 37} (2004) 841--910;
  [arXiv:math.KT/0306138].

\bibitem{Carey}
A. L. Carey, K. C. Hannabuss, V. Mathai and P. McCann, {\it
Quantum Hall Effect on the Hyperbolic Plane}, Commun. Math. Phys.
{\bf 190} (1998) 629--673; [arXiv:dg-ga/9704006].

\bibitem{Kasparov84}
G. G. Kasparov, {\it Lorentz Groups, K-Theory of Unitary
Representations and Crossed Products}, Sov. Math. Dokl. {\bf
29} (1984) 256--260.

\bibitem{Julg}
P. Julg and G. G. Kasparov, {\it Operator K-Theory for the Group
$SU(n,1)$}, J. Reine Angew. Math. {\bf 463} (1995) 99--152.

\bibitem{Cuntz}
J. Cuntz, {\it K-Theoretic Amenability for Discrete Groups}, J.
Reine Angew. Math. {\bf 344} (1983) 180--195.

\bibitem{Thurston}
W. Thurston, {\it Three-Dimensional Manifolds, Kleinian Groups and
Hyperbolic Geometry}, Bull. Amer. Math. Soc. {\bf 6} (1982)
357--381.

\bibitem{Apanasov}
B. N. Apanasov, {\sl Geometry of Discrete Groups and Manifolds}
(Nauka, Moscow, 1991).

\bibitem{Packer}
J. Packer and I. Raeburn, {\it Twisted Crossed Products of
$C^*$-Algebras}, Math. Proc. Cambridge Phil. Soc. {\bf 106} (1989)
293--311.

\bibitem{Elliott}
G. Elliott, {\it On the K-Theory of the $C^{*}$-Algebra Generated
by a Projective Representation of a Torsion-Free Discrete Group},
Monogr. Stud. Math. {\bf 17} (1984) 157--184.

\bibitem{Bellissard}
J. Bellissard, {\it K-Theory of $C^*$-Algebras in Solid State
Physics}, Lect. Notes Phys. {\bf 257} (1986)
99--156.

\bibitem{Connes}
A. Connes, {\it Noncommutative Differential Geometry}, Publ. Math.
IHES {\bf 62} (1986) 257--360.

\bibitem{Marcolli}
M. Marcolli and V. Mathai, {\it Twisted Index Theory on Good
Orbifolds I: Noncommutative Bloch Theory}, Commun.
Contemp. Math. {\bf 1} (1999) 553--587;
[arXiv:math.DG/9911102].

\bibitem{Mathai}
V. Mathai, {\it Heat Kernels and the Range of the Trace on
Completions of Twisted Group Algebras}, Contemp. Math.
{\bf 398} (2006) 321--346; [arXiv:math.KT/0606790].

\bibitem{Olsen}
K. Olsen and R. J. Szabo, {\it Constructing D-Branes from
K-Theory}, Adv. Theor. Math. Phys. {\bf 3} (1999) 889--1025;
[arXiv:hep-th/9907140].

\end{thebibliography}
\end{document}